\documentclass[journal=jpccck,manuscript=article]{achemso}
\usepackage[version=3]{mhchem} 
\usepackage[T1]{fontenc}       

\author{R. Shendrik}
\affiliation[igc]{Vinogradov Institute of geochemistry SB RAS, Favorskogo 1a, Irkutsk, Russia, 664033}
\email{r.shendrik@gmail.com}
\author{A.~Myasnikova}
\affiliation[igc]{Vinogradov Institute of geochemistry SB RAS, Favorskogo 1a, Irkutsk, Russia, 664033}
\author{A.~Shalaev}
\affiliation[igc]{Vinogradov Institute of geochemistry SB RAS, Favorskogo 1a, Irkutsk, Russia, 664033}
\author{A.~Bogdanov}
\affiliation[igc]{Vinogradov Institute of geochemistry SB RAS, Favorskogo 1a, Irkutsk, Russia, 664033}
\alsoaffiliation[irnitu] {Irkutsk National Researcher Technical University, Lermontov str. 83, Irkutsk, Russia, 664074}
\author{E.~Kaneva}
\affiliation[igc]{Vinogradov Institute of geochemistry SB RAS, Favorskogo 1a, Irkutsk, Russia, 664033}
\author{A.~Rusakov}
\affiliation[igc]{Vinogradov Institute of geochemistry SB RAS, Favorskogo 1a, Irkutsk, Russia, 664033}
\author{A.~Vasilkovskyi}
\affiliation[igc]{Vinogradov Institute of geochemistry SB RAS, Favorskogo 1a, Irkutsk, Russia, 664033}

\title[An \textsf{achemso} demo]
  {Optical and structural properties of $\mathrm{Eu^{2+}}$ doped BaBrI and BaClI crystals}

\abbreviations{VRBE,VASP,VUV}
\usepackage{amssymb}
\usepackage{latexsym}
\usepackage{textcomp}
\usepackage{amsthm}
\usepackage{amsmath}
\usepackage{subfigure}
\usepackage{hyperref}

\begin{document}

\begin{abstract}
The work is necessitated by search for new materials to detect ionizing radiation. The rare-earth ions doped with ternary alkali
earth-halide systems are promising scintillators showing high efficiency and energy resolution. Some aspects of crystal growth and
data on the structural and luminescence properties of BaBrI and BaClI doped with low concentrations of $\mathrm{Eu^{2+}}$ ions are reported. The crystals are grown by the vertical Bridgman method in sealed quartz ampoule. New crystallography data for
BaClI single crystal obtained by single crystal X-ray diffraction method are presented in this paper. Emission, excitation and optical absorption spectra as well as luminescence decay kinetics are studied under excitation by X-ray, vacuum ultraviolet and ultraviolet radiation. The energies of the first 4f-5d transition in $\mathrm{Eu^{2+}}$ and band gap of the crystals have been obtained. We have calculated the electronic band structure of the crystals using density functional theory as implemented in the \latin{Ab Initio}. Calculated band gap energies are in accord with the experimental estimates. The energy of gaps between the occupied Eu$^{2+}$ 4f level and the valence band top are predicted. In addition, positions of lanthanide energy levels in relation to valence band have been constructed using the chemical shift model.

\end{abstract}


\section{Introduction}
Eu-doped orthorhombic alkali earth halides have been recently utilized as the prospective scintillators for gamma ray detection having high light yield and energy resolution. The strontium iodide crystals doped with Eu$^{2+}$ ions demonstrate excellent properties close to the theoretical limit \cite{loef2009}. This material has been more extensively studied with optical spectroscopy methods  within the past decade. Spectroscopic data measured in ultraviolet and vacuum ultraviolet (VUV) together with \latin{Ab Initio} calculations provided the information about exciton and band gap energies \cite{pustovarov2012luminescence,pankratov2013luminescence,ogorodnikov2013luminescence,alekhin2015luminescence} and is necessary for understanding mechanism of defect formation and their role in energy transfer.

Recently the research focus is shifted to the study of mixed halide compounds due to their superior light yield ~\cite{bourret2012}. In a number of barium dihalides BaFI-BaClI-BaBrI-BaBrCl, the scintillation properties have been studied for Eu-doped $\mathrm{BaFI}$, $\mathrm{BaBrCl}$ and $\mathrm{BaBrI}$~\cite{bourret2010,gundiah2010, bizarri2011scintillation,gundiah2011structure,yan2016czochralski}. Despite their excellent properties, experimental data on optical absorption and excitation spectra in spectral region of 4f-5d and band to band transitions are scarce because single-crystals doped with high concentrations of $\mathrm{Eu^{2+}}$ ions (more than 5 mol.~\%) were used. When measuring high concentration doped samples, the inner filter effects can be observed. These include reabsorption and non uniform excitation throughout the sample. These effects dramatically change the shape of excitation spectrum. Therefore, estimation of the lowest energy of 4f-5d transitions in Eu-doped BaBrI given in \cite{gundiah2011structure, gundiah2010} is not correct. Furthermore, the experimental determination of band gap in these crystals is not possible due to high absorption related to allowed $\mathrm{4f^{6}5d^{1} \rightarrow 4f^{7}}$ transitions in Eu$^{2+}$ ions. At this moment, the energy of band gap of the mixed halide compounds is based on theoretical estimates.

We investigate luminescence, electrical and structural properties of undoped BaBrI, BaBrI-0.05 mol.\% Eu$^{2+}$ and BaClI-0.1 mol.\% Eu$^{2+}$ crystals. Absorption, excitation and emission spectra, photoluminescence decay time constants, dielectric properties and pulsed height spectra are presented. The vacuum referred binding (VRBE) energy diagram is constructed in conformity with density functional study.  It displays the electron
binding energy in the ground and excited state levels of all divalent and trivalent lanthanides ions in BaBrI and BaClI crystals.

\section{Methodology}

\subsection{Growth and Structural Characterization}

The crystals were grown by the vertical Bridgman method in sealed quartz ampoules in vacuum. The temperature gradient was about 10-15~$^{o}$C/cm, and the growth rate was 1~mm/hour. The reagents used for the growth were BaBr$_2$, BaI$_2$ and BaCl$_2$ (purity 99.9\%, Lanhit,~LTD). The stoichiometric mixtures of BaBr$_2$+BaI$_2$ and BaCl$_2$+BaI$_2$ were employed. The samples were doped with of EuBr$_3$ and EuCl$_3$, respectively. 
Since the material is hydroscopic, the batch was thoroughly dried prior to sealing the ampoule diameter 10-30~mm. 
The thermogravimetric and differential scanning calorimetry methods determined the melting point, 
the level of hydration and the possible dehydration temperatures of the charge materials prior to the crystal growth. 
The melting points for BaBrI and BaClI are about 783~$^{o}$C and 815~$^{o}$C, respectively. 

The plates about 1-2~mm of thickness and 1~cm in diameter were cut and polished in glove box for optical absorption and luminescence spectra measurements.
The rest of grains were analyzed to determine structure. For pulse height spectra measurements a sample 1x1x1~cm$^3$ of BaClI-0.1~mol.\%~Eu$^{2+}$ was cut, polished and coated with polytetrafluoroethylene (PTFE) tape to maximize the light collection efficiency.

The crystallography data of BaBrI crystal were published in paper~\cite{gundiah2011structure}. The diffraction pattern of grown BaBrI crystals was in agreement with the early published data.

We report the new data for BaClI single crystal measured by single crystal X-ray diffraction method.
Structure analysis of $\mathrm{BaClI}$ crystals was carried out using a Bruker AXS D8 VENTURE dual 
source diffractometer with a Photon 100 detector under monochromatized Mo-$K_{\alpha}$~radiation. 
Low temperature data was acquired with the crystal cooled by a Bruker Cobra nitrogen Cryostat. 
Three sets of 20 frames were utilized for initial cell determination, whereas complete data were collected by several 
$\varphi$ and $\omega$ scans with $\mathrm{0.3^{o}}$ rotation, 2~s exposure time per frame and crystal-to-detector distance 40~mm. 
The data collection strategies were optimized by the APEX2 program suite~\cite{bruker_2003}, and the reflection intensities were extracted and corrected for the Lorentz-polarization by the SAINT package~\cite{bruker_2007}. 
A semi-empirical absorption correction was applied by means of the SADABS software~\cite{sheldrik}. 
It was revealed that the studied samples crystallize in orthorhombic symmetry. 
The XPREP software assisted in the determination of the space group (Pnma) and in calculation of intensity statistics. 
Finally, the least-squares refinements were refined by the program CRYSTALS~\cite{betteridge}. 
The structures were solved with the use of the charge flipping algorithm~\cite{palatinus}, 
and the space group was confirmed by the analysis of the reconstructed electronic density. 

Scale factors, atomic positions, occupancies and atomic displacement factors were  the refined parameters. 
In preliminary anisotropic refinement the $R$ values converged to $R\approx 4$.

\begin{figure}
  \centering
    \includegraphics[width=0.5\textwidth]{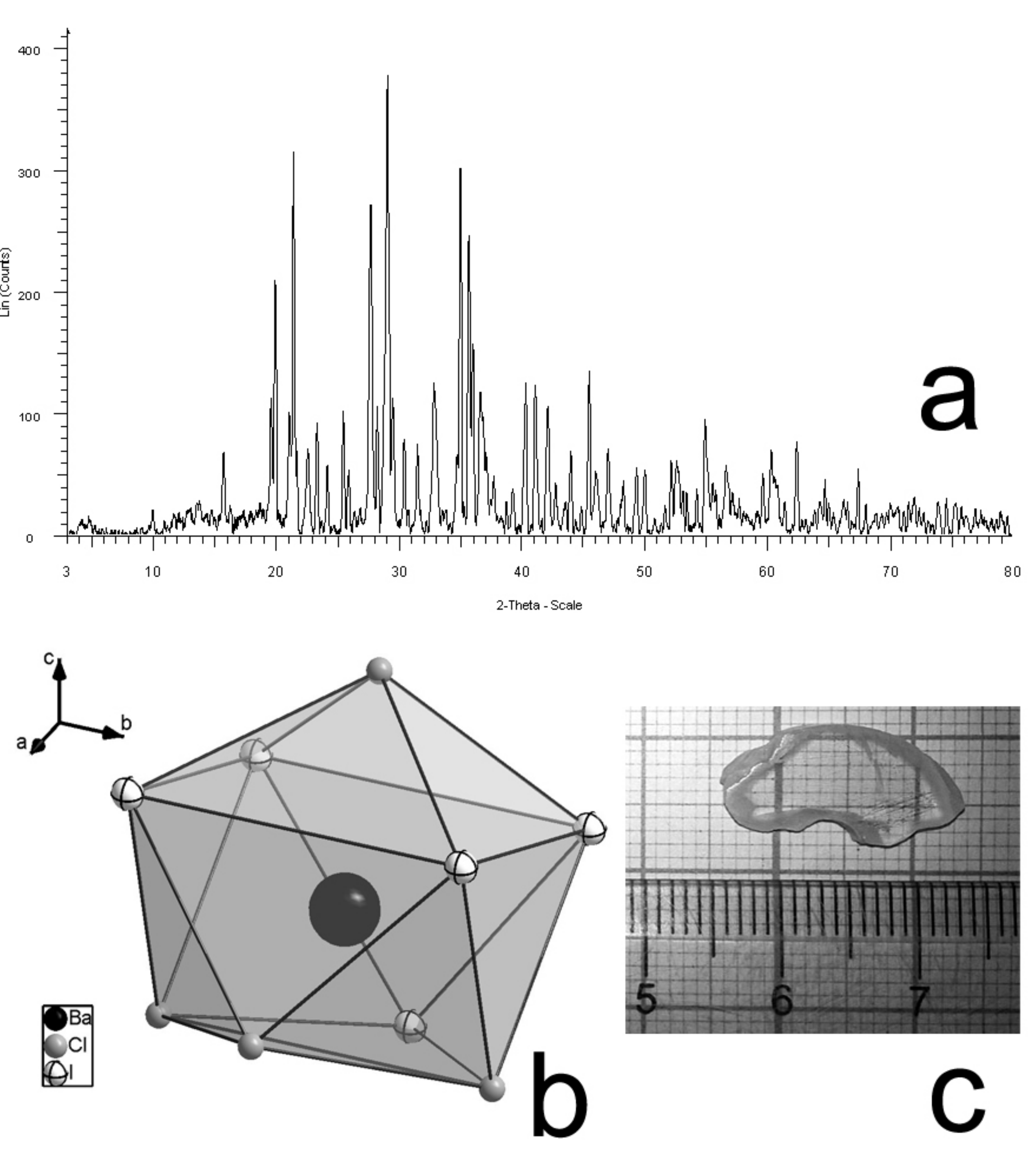} 
      \caption{(a) X-ray diffraction spectrum of powder BaClI sample. (b) Arrangement of Cl and I anions around each Ba atom. (c) Photograph of the BaClI-0.1~mol.\%~Eu$^{2+}$ crystal}\label{figure1}
\end{figure}

The obtained unit cell parameters are: $a$ = 8.4829(5), $b$= 4.9517(3), $c$ = 9.6139(5)~\AA, $V$= 403.83~\AA$^3$. The density of BaClI crystals calculated from the structure is 4.94 g/cm$^3$. In the orthorhombic structures (space group Pnma) of studied crystals Ba, Cl and I atoms 
occupy the fourfold special positions (4c) and lie on the mirror planes, perpendicular to the b axis. 
Barium atom position is coordinated by 9 anions with mean interatomic distances: 
Ba-Cl~$\sim$~3.15 and Ba-I~$\sim$~3.59~\AA (Fig.\ref{figure1}). X-ray powder diffraction data were obtained by diffractometer D8 ADVANCE Bruker in range of diffraction angles 2$\theta$ varying from 3 to 80~degrees, CuK$_\alpha$ radiation. The experimental conditions were the following: 40~kV, 40~mA, time per step - 1~s and step size - 0.02$^{o}$, Goubel mirror. The XRD pattern of the sample is shown in Fig.\ref{figure1} and, in general, it is similar to that obtained in Ref.~\cite{lenus2002luminescence}. In contrast to our data, the space group Pbam was reported for BaClI in Ref.~\cite{lenus2002luminescence}.

\subsection{Optical and dielectrical characteristic measurements}
The optical absorption spectra were obtained by a Perkin-Elmer Lambda 950 UV/VIS/NIR spectrophotometer at 300 K; 
photoluminescence (PL) was measured in vacuum cold-finger cryostat; 
The spectra were detected with a MDR2 and SDL1 (LOMO) grating monochromator, a photomodule Hamamatsu H6780-04 (185-850 nm), and a photon-counter unit. 
The luminescence spectra were corrected for spectral response of detection channel. 
The photoluminescence excitation (PLE) spectra were measured with a grating monochromators MDR2 and 200~W xenon arc lamp for direct 4f-5d 
excitation and vacuum monochromator VM-2 (LOMO) and Hamamatsu deuterium lamp L7292 for measurements in VUV spectral region. 
The excitation spectra were corrected for the varying intensity of exciting light due. 
Photoluminescence decay curves were registred by an oscilloscope Rigol 1202 under pulse nitrogen laser excitation with impulse duration about 10 ns. 
The X-ray excited luminescence was performed using an X-ray tube operating at 50 kV and 1 mA.

The measurements of pulse height spectra were carried out with a photomultiplier tube Enterprises 9814QSB. 
The PMT was operated with a CSN638C1 negative polarity voltage chain. The focusing system of 46 mm active diameter 
was assumed to be 100\% photoelectron collection efficiency in the center of the photocathode. 
The sample was irradiated with gamma rays from a monoenergetic $\gamma$-ray source of $^{137}$Cs (662~KeV). 
A homemade preamplifier and an Ortec 570 amplifier were used to obtain pulse height spectra. 
The samples are optically coupled to the window of PMT using mineral oil.

The dielectric constant of the BaBrI crystal was measured using immitance (RLC) meter E7-20 manufactured by MNIPI. 
The measurements of capacitance and dielectric losses were performed in frequency range from 25~Hz to 1~MHz. 
The silver paint (kontaktol "Kettler") was employed as the electrode contact material. The pad area was about 60 $mm^2$ and the sample thickness was about 1~mm. 
To prevent surface degradation the dieletric measurements of polished samples were made in the glove box . 

\subsection{Calculation details}

\latin{Ab Initio} calculations of  BaClI crystal doped with $\mathrm{Eu^{2+}}$ were carried out within 
density functional theory (DFT) using VASP (Vienna \latin{Ab Initio} Simulation Package) computer code~\cite{vasp}. 
The calculations were performed with HPC clusters "Academician V.M.~Matrosov" ~\cite{matrosov} 
and "Academician A.M.~Fock"~\cite{fock}. Using the unit sell parameters from the X-ray diffraction we 
constructed the 2\texttimes 2\texttimes 1 (48~atoms) supercell, 
in which one of $\mathrm{Ba^{2+}}$ ions was replaced by $\mathrm{Eu^{2+}}$.

The spin-polarized calculations were carried out within the framework of the generalized gradient approximation (GGA) 
with the exchange-correlation potential PBE~\cite{pbe}. 
Integration within the Brillouin zone was performed on a $\mathrm{\Gamma}$-centered grid of 8 irreducible \textit{k} points.
Geometry optimization was performed with fixed cell dimensions. 
The convergence was achieved if the difference in total energy between the two iterations was less than $\mathrm{10^{-6}~eV}$.

\section{Results and discussion}

\subsection{Eu$^{2+}$ luminescence}
\begin{figure}[t!]
\centering
\includegraphics[width=0.7\textwidth]{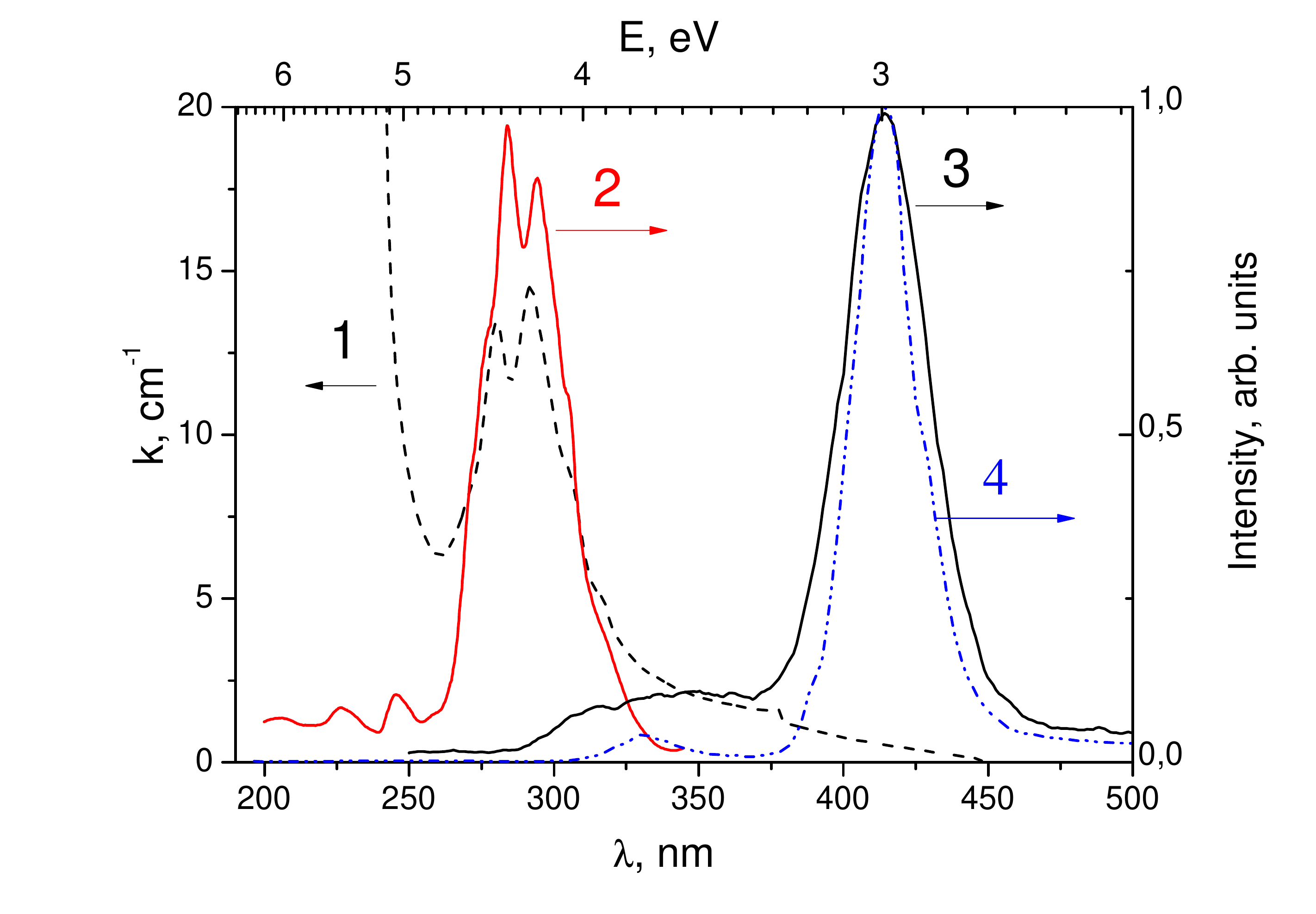}
\caption{Optical absorption (dashed curve 1), excitation (curve 2) monitored at 415 nm, 
and emission spectra under intracenter 4f-5d excitation at 290 nm (curve 3) and x-ray excitation (dot curve 4) of BaBrI crystals doped 
with 0.05 mol.\% Eu$^{2+}$ ions. }
\label{lumin_babri}
\end{figure}

\begin{figure}[t!]
\centering
\includegraphics[width=0.7\textwidth]{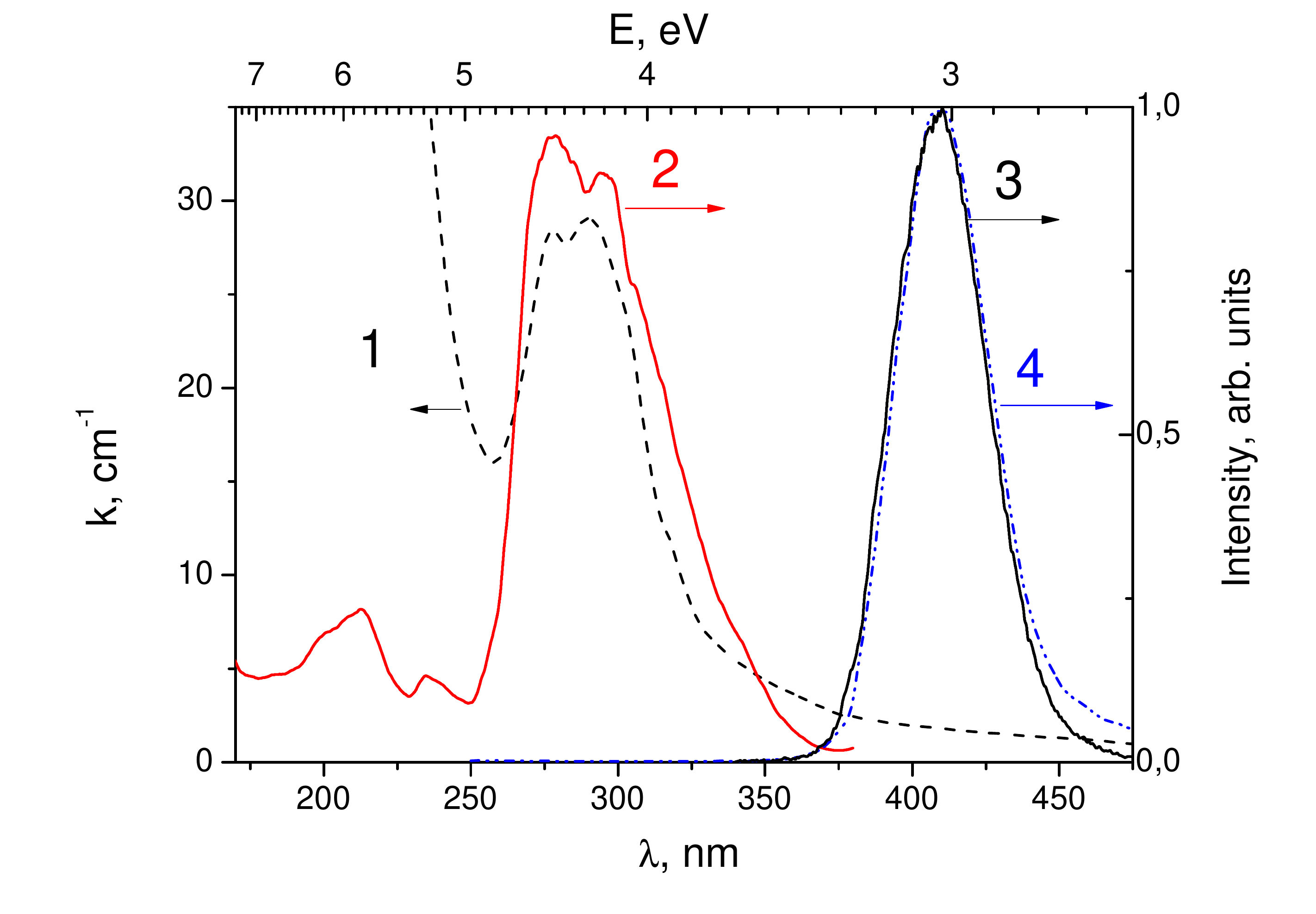}
\caption{Optical absorption (dashed curve 1), excitation (curve 2) monitored at 410 nm, 
and emission spectra under intracenter 4f-5d excitation at 290 nm (curve 3) and x-ray excitation (dot curve 4) of BaClI crystals doped with 
0.1 mol.\% Eu$^{2+}$ ions. }
\label{lumin_bacli}
\end{figure}

Figures~\ref{lumin_babri}~and~\ref{lumin_bacli} exhibit the absorption spectra for Eu$^{2+}$ ions in BaBrI and BaClI crystals measured at room temperature and the relative emission and excitation spectra measured at 80 K. 
Strong absorption bands are observed from 4f$^7$ ground state to 4f$^6$5d$^1$ states in both figures (curves~1). 
The maxima of peaks are 280 (4.43~eV) and 292 nm (4.25~eV) for BaBrI  crystal and 278 (4.46~eV) and 290 nm (4.25~eV) for BaClI crystal. 

The excitation wavelength for the emission is 290~nm. At this wavelength, the optical penetration depth is less than 1~mm for all the samples.  
The emission spectra, which result from 5d-4f transitions with peaks at 415~nm (2.99~eV) in BaBrI-Eu and 410~nm (3.02 eV) in BaClI-Eu 
(curves 3 in Fig.~\ref{lumin_babri} and ~\ref{lumin_bacli}), agree well with previously published data \cite{gundiah2011structure, lenus2002luminescence}.

The excitation spectra of 5d-4f emission of BaBrI-Eu and BaClI-Eu crystals monitored at 415 and 410~nm (curves 2 in Fig.~\ref{lumin_babri} and ~\ref{lumin_bacli}) are well agreed with the optical absorption spectra.

The ground state of Eu$^{2+}$ ions includes seven 4f electrons, which, according to the Hund's rules, give rise to the ground state $^8$S$_{7/2}$. 
The bands consist of two clearly distinguishable peaks and slightly expressed characteristic "staircase" structure. In BaBrI and BaClI crystals the site symmetry of cations is D$_{2h}$ as found in structure analisis. In D$_{2h}$ symmetry the degeneracy of the t$_{2g}$ is lifted and the t$_{2g}$ level splits into three levels. Excitation and absorption spectra indicate two bands corresponding to a t$_{2g}$ level splitting into three levels, where degeneracy of two of them is weak.
The characteristic "staircase" structure was originally explained by transitions from the $^8$S$_{7/2}$ to the seven $^7$F$_J$ multiplets (J = 0-6) 
of the excited 4f$^6$($^7$F$_J$)5d$^1$ configuration \cite{dorenbos2003energy}.

It is feasible to the lowest energy of 4f-5d transtion ($\lambda_{abs}$) from the absorption spectra for BaBrI-Eu and BaClI-Eu. 
According to P. Dorenbos \cite{dorenbos2003energy}, this value pertains to the first step of the characteristic "staircase" structure 
in 4f-5d absorption and excitation spectra of Eu$^{2+}$ corresponding to the zero phonon-line in emission spectrum, 
such as CaF$_2$ doped Eu$^{2+}$ ions \cite{kobayasi1980fluorescence}. 
Usually "staircase" structure and vibronic structure of Eu$^{2+}$ emission band are not resolved. 
Therefore, the $\lambda_{abs}$ value is estimated on the energy of low-energy side, where the band has risen to 15--20\% of the maximum of the "staircase". 
In this method some level of arbitrariness may introduce an error. 
To keep errors small, the data on samples with low Eu concentration were preferably used. 
In previous works \cite{chaudhly}, the estimation of $\lambda_{abs}$ 397~nm  is based on the measurement of crystals 
doped with high concentrations of Eu$^{2+}$ ions (more than 5 mol.\%). 
Thus, excitation spectra could not be measured correctly due to self-absorption and $\lambda_{abs}$ value contains a large error. 
From absorption spectra of BaBrI-0.05 mol.\% Eu$^{2+}$ and BaClI-0.1 mol.\% Eu$^{2+}$ we found $\lambda_{abs}$(BaBrI)=376~nm for 
BaBrI-Eu$^{2+}$ and  $\lambda_{abs}$(BaClI)=373~nm. 
The lowest energies of 4f-5d transition in Eu$^{2+}$ ion are 3.29~eV for BaBrI and 3.32~eV for BaClI. 

At all temperatures the investigated crystals show only strong 5d-4f luminescence and no 4f-4f emission. 
At room temperature Eu$^{2+}$ compounds can exhibit broad and strong fluorescence resulting from the 5d-4f transitions, as well as sharp line emission, which has been assigned to 4f-4f transitions from $^6$P$_{7/2}$ to $^8$S$_{7/2}$ terms. 
The presence of both 5d-4f and 4f-4f transitions indicates the proximity of the lowest excited 5d and the 4f$^7$ ($^6$P$_{7/2}$) states. 
The relative positions of the 5d and  $^6$P$_{7/2}$ levels change in different hosts. 
Strong 5d-4f band emission is observed when the lowest 5d state is significantly lower than the  $^6$P$_{7/2}$ level, 
f-f sharp line emission appears when the  reverse is true. Both 5d-4f and f-f bands in luminescence spectrum are obtained when the energy of the levels are close. 
Decreasing temperature causes to redistribution of intensities 4f-4f and 5d-4f luminescence lines. 
Increase of 4f-4f luminescence and the corresponding reduction of 5d-4f luminescence at low temperatures take place, when the lowest 5d state 
lies slightly above $^6$P$_{7/2}$ terms of 4f state. 
That type of luminescence was observed in Eu$^{2+}$ doped BaFCl and SrFCl crystals \cite{kobayasi1980fluorescence}. 
Energy of $^6$P$_{7/2}$ level in relation to the ground $^8$S$_{7/2}$ level is about 3.49~eV. 
Thus, it can be concluded that the energy 5d state in relation to $^8$S$_{7/2}$ level is significantly lower than 3.49~eV. 
Therefore, our estimation of the lowest 4f-5d energies should be reliable.

\begin{figure}[t!]
\centering
\includegraphics[width=0.7\textwidth]{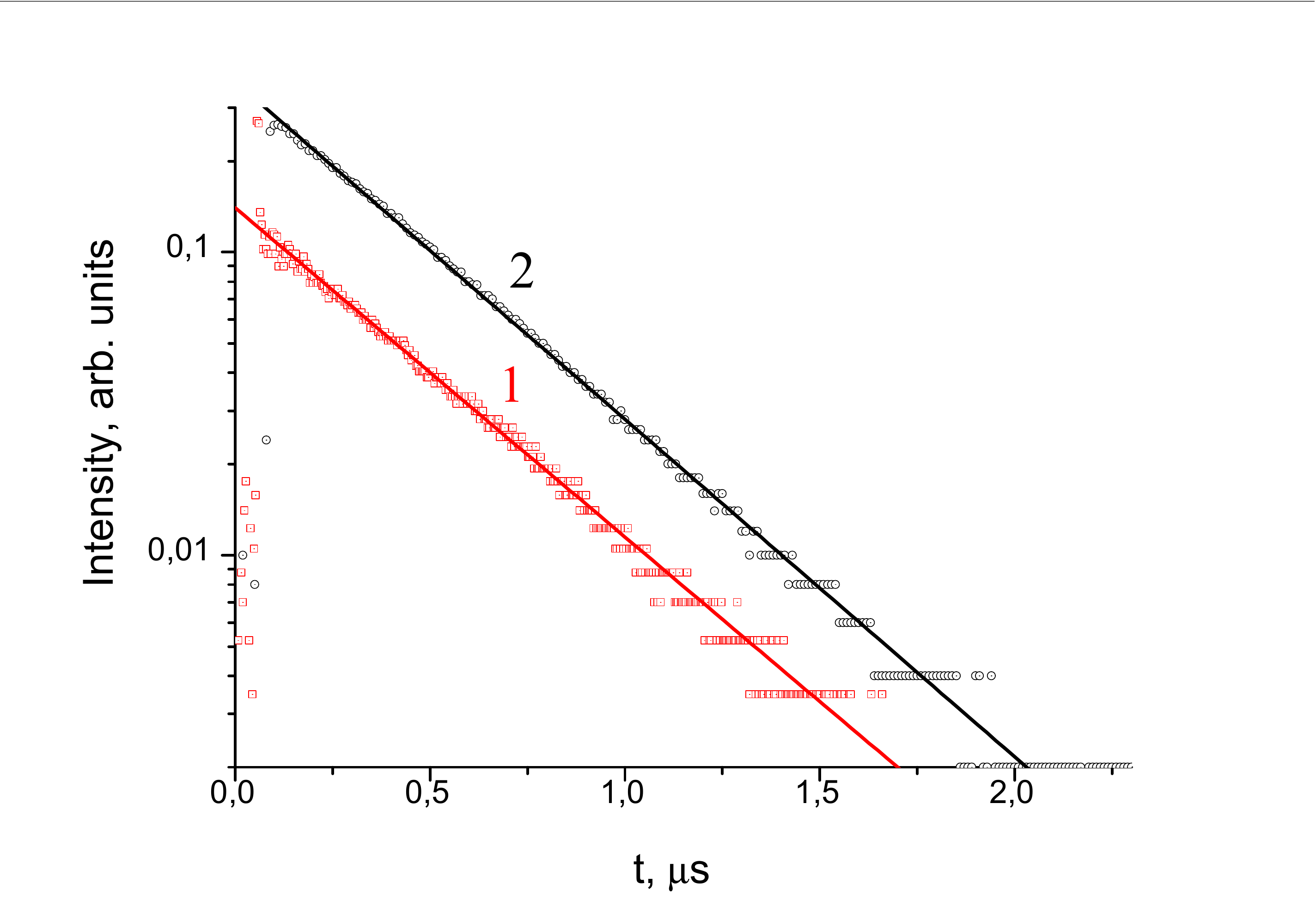}
\caption{Luminescence decay curves of BaClI-0.1 mol.\% Eu$^{2+}$ (curve 1) and BaBrI-0.05 mol.\% Eu$^{2+}$ (curve 2) excited at 337 nm}
\label{decays}
\end{figure}

Luminescence decay curves measured at the maximum of the emission peak under nitrogen laser excitation at 337 nm are monoexponential in shape 
(Fig.~\ref{decays}). The measured time constants of the photoluminescence decay kinetics were equal to $\tau$=390~ns for BaClI-Eu 
(Fig.~\ref{decays} curve 1) and $\tau$=400~ns for BaBrI-Eu (Fig.~\ref{decays}) crystals. 
The values are quite similar to the time-constants obtained with other types of excitations \cite{gundiah2011structure,  bizarri2011scintillation}. 

BaBrI and BaClI crystals doped with Eu$^{2+}$ ions demonstrate bright luminescence under x-ray and gamma-ray excitation. 
The spectra of x-ray excited luminescence  are given in curves 4 in Fig.~\ref{lumin_babri} and ~\ref{lumin_bacli}. 
The spectra are in agreement with photoluminescence spectra. 
The X-ray excited luminescence output measured by integral intensity is compared with the one of CaF$_2$-Eu crystal. 
Light output of CaF$_2$-Eu crystals is approximately 21000 photons/MeV \cite{shendrik2013scintillation, shendrik2014absolute}. 
Therefore, the light output of measured samples can be estimated. BaBrI doped with 0.05 mol.\% Eu has light output about 25000~photons/MeV. 
Light output of BaClI doped with 0.1 mol.\% Eu$^{2+}$ is estimated about 30000~photons/MeV. 
In artciles \cite{bourret2010, BourretCourchesne} the authors obtained the best light output in BaBrI doped with 5--7~\% Eu$^{2+}$ ions. 
Therefore, we can expect to increase light output at higher than 0.1~mol.~\% concentrations of Eu$^{2+}$ ions. 

\begin{figure}[t!]
\centering
\includegraphics[width=0.7\textwidth]{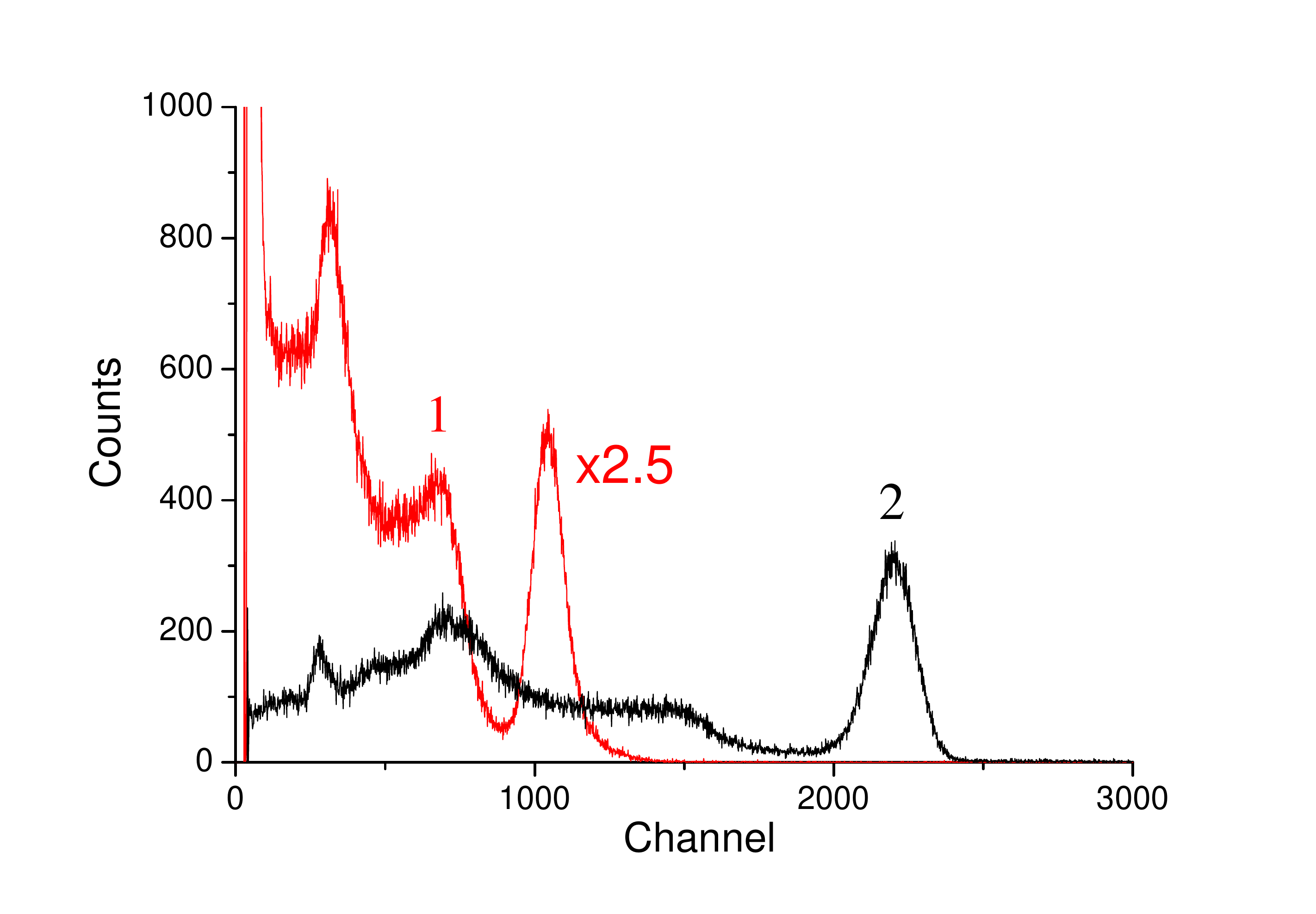}
\caption{Pulsed-height spectra of BaClI-0.1 mol.\% Eu$^{2+}$ and NaI-Tl (light output is about 41000 photons/MeV \cite{shendrik2014absolute}) }
\label{pulsed-height}
\end{figure}

Light output of BaClI-Eu calculated from the pulsed height spectrum (Fig.~\ref{pulsed-height}) is about 9000~photons/MeV and an energy resolution is about 11\%. 
This value is much lower than the one obtained from the x-ray excited luminescence spectra.  
The spread is attributed to various factors. 
The first one is the lower quality of large crystal than the small sample used in the x-ray luminescence spectra measurements. 
The next reason is in possible presence of intensive slow components in Eu$^{2+}$ x-ray and gamma excited luminescence similar 
to \cite{shendrik2013scintillation, shendrik2012energy}.

\subsection{Exciton emission and band gap}

\begin{figure}[t!]
\centering
\includegraphics[width=0.7\textwidth]{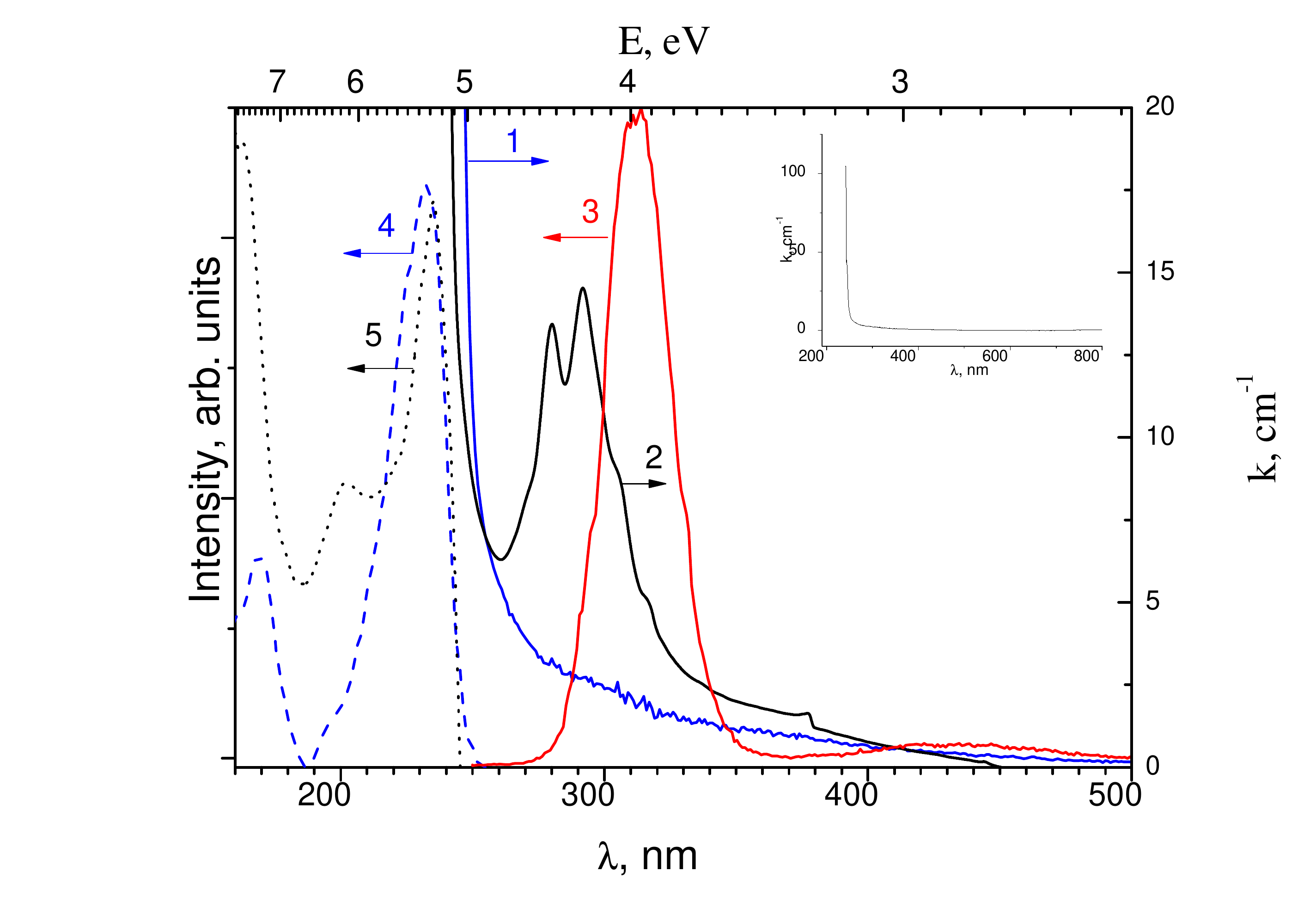}
\caption{Optical absorption spectra of undoped (curve 1) and 0.05 mol.\% Eu$^{2+}$ doped BaBrI. 
The inset contains absorption spectra of undoped BaBrI. Emission spectrum of pure BaBrI measured at 78~K corresponds to curve 3 under X-ray and interband (at about 161~nm) excitations. 
Dashed curve 4 and dotted curve 5 are excitation spectra at 320 nm emission wavelength in the undoped and Eu-doped crystals, respectively.}
\label{excit_babri}
\end{figure}

\begin{figure}[t!]
\centering
\includegraphics[width=0.7\textwidth]{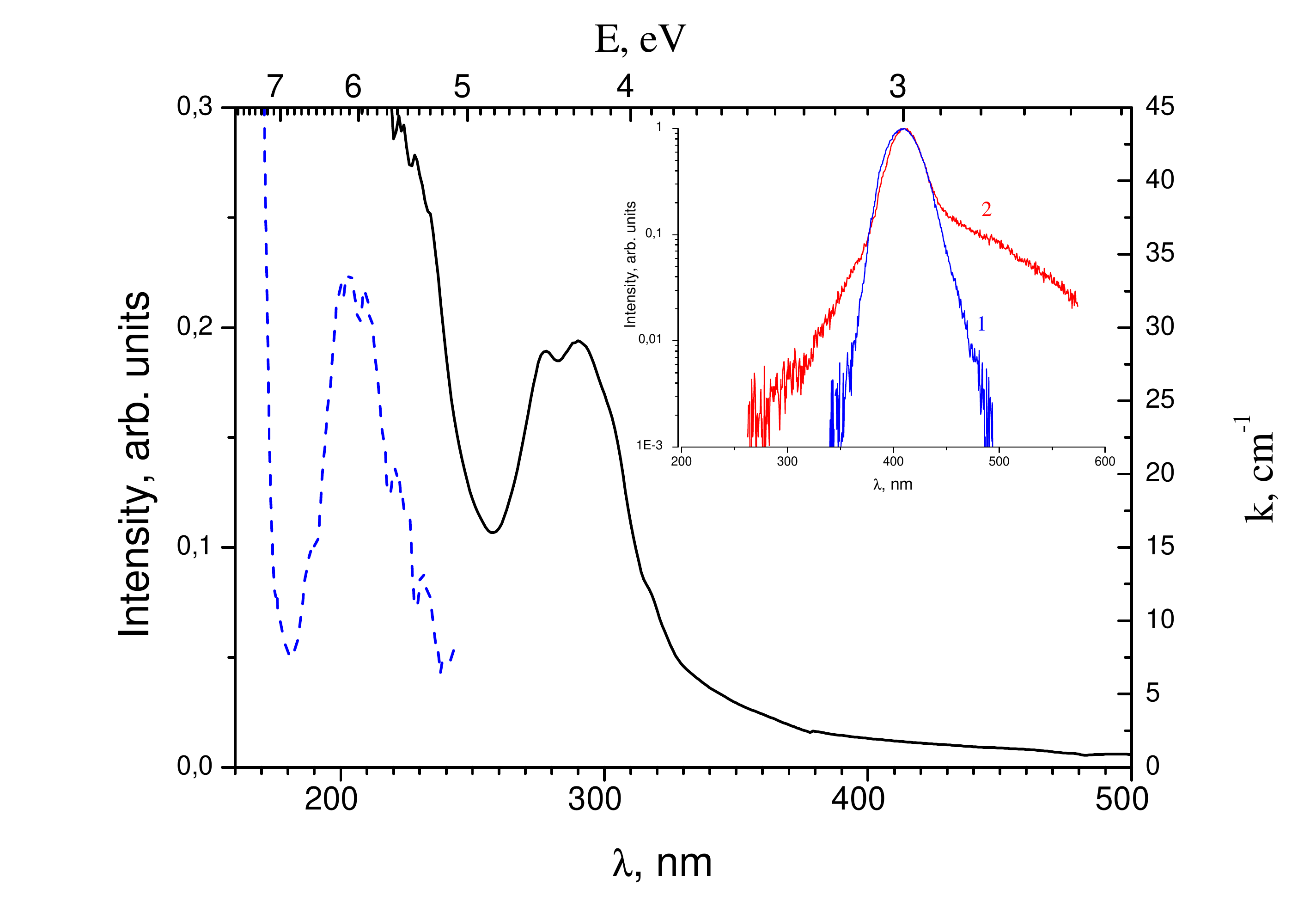}
\caption{Excitation spectrum of at 320 nm emission wavelength (dashed curve) measured at 77 K in comparison with absorption spectrum of 
BaClI-0.1 mol.\% Eu$^{2+}$ (solid curve). The inset displays emission spectra under 290 nm (curve 1) and interband (curve 2) excitations}
\label{excit_bacli}
\end{figure}

In the X-ray excited luminescence spectra of the Eu-doped samples we observe a low intensity peak at higher energy (about 3.8--4~eV) 
together with intense Eu attributed luminescence. 
In nominally undoped BaBrI crystals this luminescence dominates, and its intensity increases at low temperatures (Fig.~\ref{excit_babri}, curve 1). 
Under VUV excitation we also observe the same emission. 
The instensity of luminescence decreases with concentration of Eu$^{2+}$ ions and it is almost absent at 0.1~mol.\% and higher Eu$^{2+}$ concentrations. 
In BaClI we also obtain luminescence band in the 300--320~nm spectral region. 
The inset of Fig.~\ref{excit_bacli} contains the luminescence spectra under excitation to 4f-5d Eu$^{2+}$ band (curve 2) and VUV excitation at about 165~nm. The spectra look different from each other. Under VUV excitation the additional relatively weak bands apear at about 300--320~nm and 460--520~nm. The excitation spectra of the luminescence centred at 320~nm in undoped and doped with 0.05~mol.\% Eu$^{2+}$ BaBrI crystals are shown in Fig.~\ref{excit_babri} in comparison with optical absorption spectra. 
In all samples the most efficient photoluminescence excitation ranges from about 5.1 to 5.7~eV and lies within interband absorption spectrum.

The excitation spectrum of 300--320 nm emission band in BaClI-0.1 mol.\% Eu crystal is provided in figure~\ref{excit_bacli}. 
Similarly to BaBrI crystals a narrow peak in fundamental absorption region is found. 
The observed luminescence is attributed to self-trapped excitons (STE). 
As it was first shown by Hayes and confirmed by Song and Williams \cite{song2013self, beaumont1970investigation} 
the STE in alkaline-earth fluorides consists of molecular ion similar to H-centre (hole on interstitial fluorine) 
and an F-centre-like part (electron trapped fluorine vacancy). 
Radzhabov pointed out that STE in barium dihalides has the configuration similar to the excitons in alkali-earth fluorides \cite{radzhabov1994exciton}. 
In SrI$_2$ crystal having close band gap energy the alike emission was also ascribed to self-trapped exciton 
emission \cite{ogorodnikov2013luminescence, pankratov2013luminescence}. 
The STE luminescence quenching with Eu$^{2+}$ concentration is due to the exciton emission and 4f-5d Eu$^{2+}$ absorption spectra overlapping. 
This fact makes possible a resonant transfer from exciton to Eu$^{2+}$ ions.

Emission peaked at 480~nm and excited at about 245--250~nm can be attributed to oxygen-vacancy centres. 
This oxygen related emitting center has also been identified and discussed in the other alkali earth metal halides, such as BaFCl and 
BaFBr \cite{radzhabov1995optical} and in SrI$_2$ \cite{pustovarov2012luminescence}, \cite{pankratov2013luminescence}. 
This luminescence has been previously suggested to be STE luminescence \cite{bizarri2011scintillation}. 
Pustovarov \cite{pustovarov2012luminescence} concluded that a significant part of the oxygen contamination comes from the surface hydrate 
reactions in strontium iodide and similar mechanism can take place in the investigated crystals. 
The oxygen luminescence in contrast to the exciton emission is quenched at low temperatures similarly to the other barium dihalides \cite{radzhabov1994time,radzhabov1995optical}.

To estimate exction binding energy we need to measure the dielectric constant of the crystals. 
The dieletric constant ($\varepsilon^{'}$) is calculated using measured capacitance at different frequencies. 
The value of $\varepsilon^{'}$ at 1~MHz is about 9.93$\pm$0.2.

Assuming hydrogenlike energy levels for a simple exciton model the exciton binding energy is plotted as function of 1s hydrogen-like wave function, then its Bohr radius $a^*_0$. 
Let assume that the variation of $m=m^*$ is relatively small in the considered ionic crystals (BaFBr, BaFI, KBr, RbI, NaI, CaF$_2$, SrF$_2$, BaF$_2$). Therefore, considering the Ref.~\cite{dvorak2013origin} we can suppose that exciton binding energy:
\begin{equation}
\label{bind_exciton}
 E_b=\frac{e^2}{\varepsilon^{'}a_{0}^{*}}=\frac{e^2}{(m^{*}/m)a_0(\varepsilon^{'})^{2}}=\frac{E_{0}}{(m^{*}/m)(\varepsilon^{'})^{2}},
\end{equation} 
where $m$($m^{*}$) is electron (effective reduced) mass, $\varepsilon^{'}$ is the dielectric constant of material, $a_0$ is the Bohr radius for the hydrogen atom, $E_0=13.6$~eV is the ionization energy of hydrogen atom, and $e$ is the electron charge.

\begin{figure}[t!]
\centering
\includegraphics[width=0.7\textwidth]{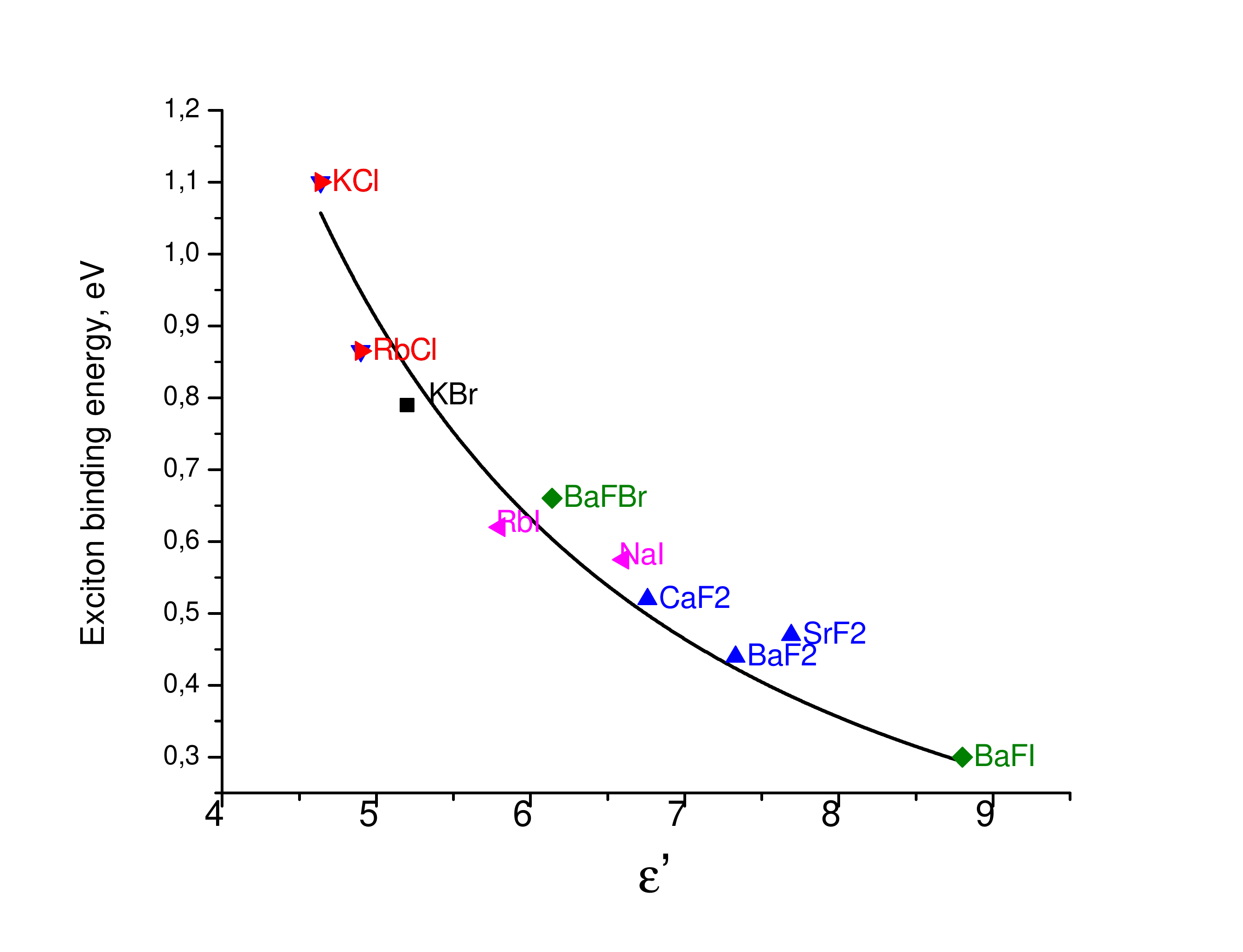}
\caption{Dependence of an exciton binding energy ($E_b$) on dielectric constant for various ionic crystals. 
Solid line fits the curve following eq.~\eqref{bind_exciton} $E_b=22.75/(\varepsilon^{'})^2$}
\label{exc_binding}
\end{figure}

To estimate exciton binding energy in BaBrI and BaClI we plot exciton binding energies versus dielectric constants for a row of known materials. 
Exciton binding energies for KBr, RbI and NaI are known from Refs.~\cite{ramamurti1966intrinsic, williams1974excited, onodera1967excitons}. 
The binding energies of CaF$_2$, SrF$_2$, and BaF$_2$ were obtained in \cite{tomiki1969optical}.
The energies for BaFI and BaFBr crystals are estimated from measurements in Ref.~\cite{nicklaus1979optical}. 
They are 0.66 and 0.32 eV for BaFBr and BaFI, respectivelly. The dielectric constants for BaFBr and BaFI are equal to 6.14 and 8.8 
respectively following by \cite{ayachour1991electrical}.  
It is evident in Fig.~\ref{exc_binding} that model quite correctly describes delocalized excitons created under 
interband excitation at the initial time in ionic crystals. In this model, the $n$-th level ($n$=1,2,3, . . . ) 
of the exciton energy $E_x$ is expressed as \cite{tsujibayashi1999resonant}:

\begin{equation}
\label{hydrogen_exciton}
E_x=E_g-\frac{E_b}{n^2},
\end{equation} 
where $E_g$ is the band gap of crystal and $E_b$ is exciton binding energy. 
Since the model above gives a good agreement between dielectric constants and binding energies for diverse crystals we can conclude 
that the primary process is excitation of conduction electrons and valence holes or directly free exciton. 
The next step is rapid localisation into a self-trapped exciton (STE). It is clear that this model can not be applied to the STE. 

Considering the spread of values for the crystals it is possible to estimate exciton binding energy in BaBrI crystal as 0.23$\pm$0.02 eV. 
Exciton energy ($E_x$) corresponds to the exciton peak in Fig. \ref{excit_babri}, \ref{excit_bacli}. 
It can be determined following the procedure proposed  by Ref.~\cite{belsky1999luminescence}. The exciton energies are about 5.35$\pm$0.15~eV for BaBrI and 6.0$\pm$0.4 eV for BaClI. 
Therefore, the band gap of the crystal is obtained by adding 1s exciton energy to the exciton binding energy (eq. (\ref{hydrogen_exciton})). 
The one is $E_g$=5.58$\pm$0.17~eV for BaBrI. 
This value is smaller than the one estimated from Dorenbos empirical rule $E_g=1.08\cdot E_x$ \cite{dorenbos2005eu3+}.  
The 1.08 proportionality factor was determined as the average of the limited available data. 
The error on the location of the conduction band depends on the type of anions in material. For most of iodides this rule gives less than 68\% 
prediction interval \cite{joos2015energy}.

We did not measure dielectric constant for BaClI, nevertheless we suppose that the dielectric constant value lies between 8.8 ($\varepsilon^{'}$ (BaFI)) 
and 9.91 ($\varepsilon^{'}$ (BaBrI)). 
We use the value 9.35, that is the mean in the interval 8.8 and 9.9, as dielectric constant of BaClI for exciton binding energy estimation. 
Therefore, exciton binding energy in BaClI is about 0.26$\pm$0.05 eV and band gap of BaClI is about 6.26$\pm$0.3 eV. 

\subsection{Calculation data}
The ground state was calculated in order to estimate the location of $\mathrm{Eu^{2+}}$ 5d and 4f levels 
relative to the valence band maximum (VBM) and conduction band minimum (CBM) respectively.
The calculations were carried out assuming that 
4f and 5d levels of europium ion must be located in the band gap of the crystal.

The ground state of $\mathrm{Eu^{2+}}$ ion configuration $\mathrm{[Xe]4f^{7}}$ is characterized by a half-filled 4f shell. 
In Refs.~\cite{chaudhly,canning} it was shown that a correct description of the 4f electrons requires 
a corrections of the effective on-site Coulomb interaction between f-electrons (characterized by the Hubbard $U$ value).
Therefore to correct the position of 4f levels the Dudarev's approximation PBE+U ~\cite{dudarev} was applied, 
in which only a difference $\mathrm{U_{eff}=(U - J)}$ is meaningfull without the individual parameters $U$ and $J$. 
In accordance with literature $\mathrm{U_{eff}}$ for $\mathrm{Eu^{2+}}$ doped wide gap materials should be~$\mathrm{\geq 4}$~\cite{holsa,shi}. 
However, authors of Ref.~\cite{chaudhly} showed that the most correct $\mathrm{Eu^{2+}}$ 4f levels described with $\mathrm{U_{eff}}$ from 2.2 to 2.5.

Some methodological calculations of $\mathrm{BaBrI:Eu^{2+}}$ crystals were performed to identify $\mathrm{U_{eff}}$ value for the method. 
The results of the calculations are almost identical to the ones obtained in Ref.~\cite{chaudhly}. 
The value $\mathrm{U_{eff}}$ was chosen to be 2.5, which gave a good agreement with the data reported in Ref.~\cite{chaudhly} 
(4f-VBM = 1.4 eV) for $\mathrm{BaBrI:Eu^{2+}}$). 
The calculated 4f-VBM gap for $\mathrm{BaClI:Eu^{2+}}$ was 1.5~eV at $\mathrm{U_{eff}=2.5}$.

The band gap was estimated both by the PBE, and by the $\mathrm{G_{0}W_{0}}$ approximations~\cite{shishkin,fuchs}. 
It is known that the using of density functional calculations with PBE potential in semiconductors and dielecrics 
leads to a delocalized electron states and, consequently, to underestimate the energy of the band gap~\cite{lee}.
However, the $\mathrm{G_{0}W_{0}}$ method gives the value of band gap in ionic crystals, comparable with 
the experimental data~\cite{chaudhly,canning,lee}, which is also confirmed by the calculations.
The results of the calculations are presented in Table~\ref{all-data}).
Finally, we can estimate the energy of 5d-CBM transition using the calculated band gap and 4f-VBM energy, and experimental data 
of the first $\mathrm{4f\rightarrow5d}$ transtion.
The estimated 5d-CBM values were 0.65 and 0.75~eV for $\mathrm{BaBrI:Eu^{2+}}$ and $\mathrm{BaClI:Eu^{2+}}$ and agreed well with the data obtaned in Ref.~\cite{chaudhly}.

\begin{figure}[t!]
  \centering
    \includegraphics[width=0.7\textwidth]{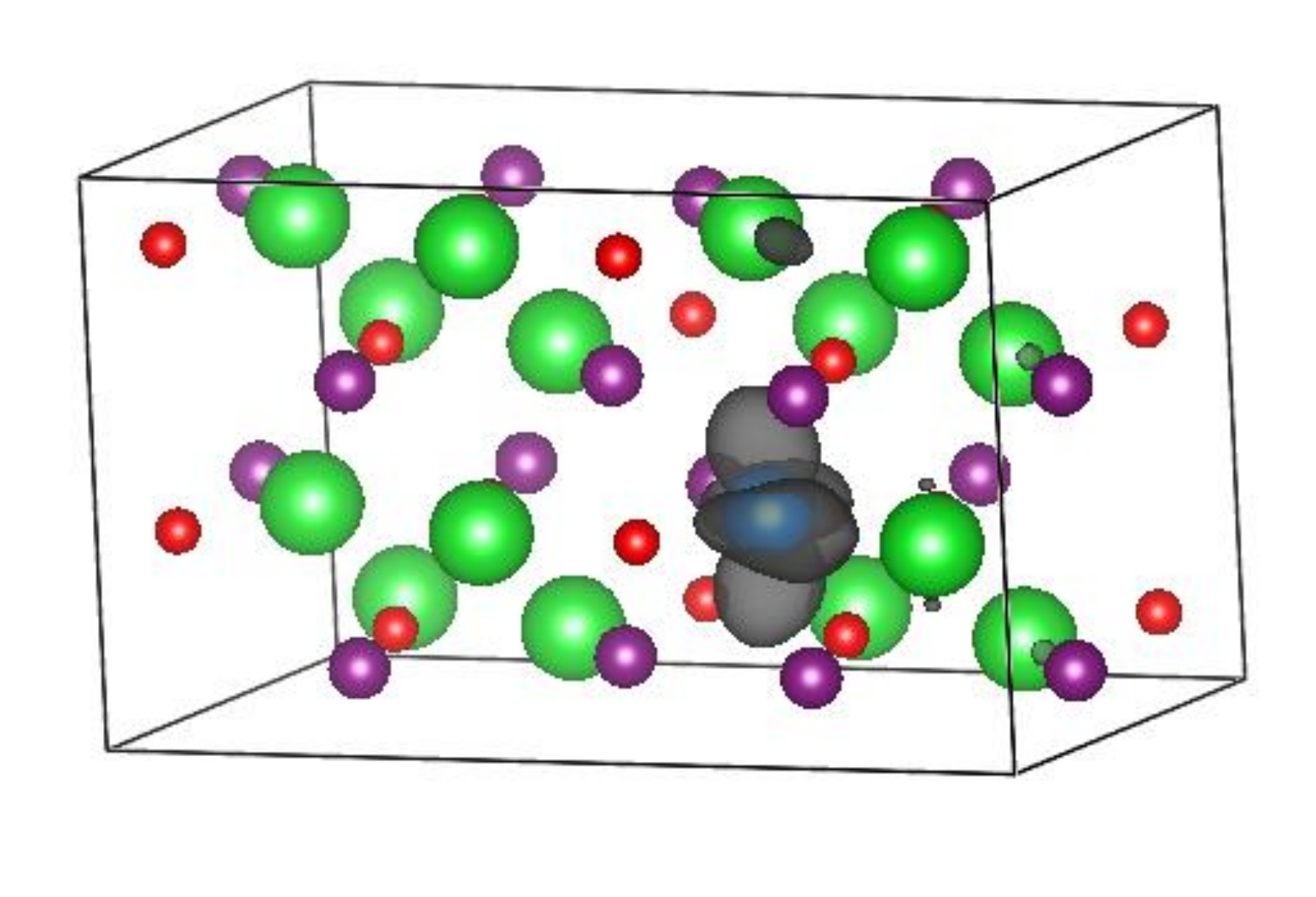} 
      \caption{Isosurface of the electron density for the $\mathrm{Eu^{2+}}$ first excited state of $\mathrm{BaClI:Eu^{2+}}$.
 Eu atom is shown in blue, Ba - in green, Cl~- in mangenta, I~- in red.}\label{figure2}
\end{figure}

The excited state of $\mathrm{[Xe]4f^{6}5d^{1}}$ configuration of $\mathrm{Eu^{2+}}$ ion was performed by setting the occupancy 
of the highest 4f state to zero. 
The isosurface of electron density for the excited state of $\mathrm{Eu^{2+}}$-doped BaClI is shown in Fig.~\ref{figure2}.
The $\mathrm{5d^{1}}$  excited state is almost completely localized on the Eu ion for both investigated crystals.

\begin{table*}
\begin{center}
 \caption{Calculated band gaps and relative 4f and 5d levels for $\mathrm{Eu^{2+}}$-doped BaBrI and BaClI crystals. Energies are given in eV.}
\label{all-data}
  \begin{tabular}{|l|c|c|c|c|c|c|}
  \hline
Crystal & \multicolumn{3}{|c|}{Band gap} & $\mathrm{Eu^{2+}}$, & $\mathrm{Eu^{2+}}$, &  $\mathrm{Eu^{2+}}$, \\
 & PBE & $\mathrm{G_{0}W_{0}}$ & Exp. & 4f - VBM & $\mathrm{4f\rightarrow 5d}$ & 5d - CBM \\
 \hline
BaBrI & 3.49 & 5.34 & 5.58$\pm$0.17 & 1.4 & 3.29 & $\sim 0.65$ \\
\hline
BaClI & 3.71 & 5.57 & 6.26$\pm$0.3 & 1.5 & 3.32 & $\sim 0.75$\\
\hline
  \end{tabular}
\end{center}
\end{table*}

\subsection{ Vacuum referred binding energy diagram of the lanthanide ions in BaBrI and BaClI}

\begin{figure*}[t!]
\centering
\includegraphics[width=1.07\textwidth]{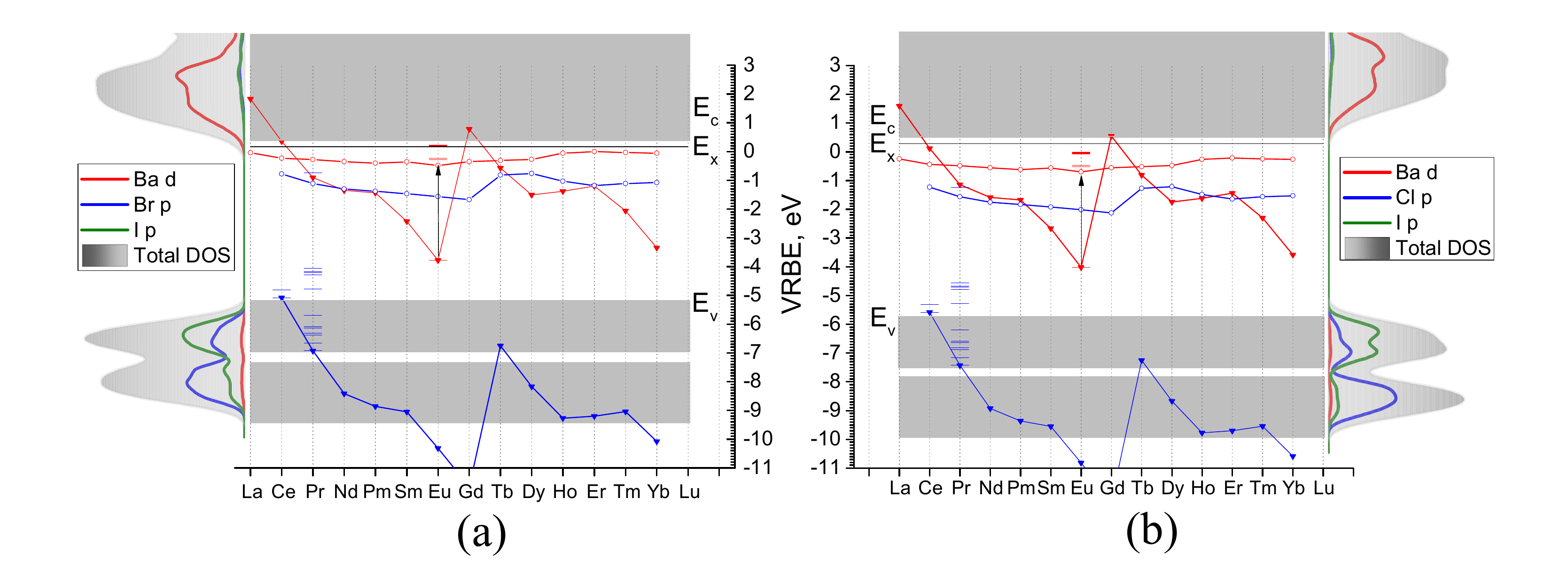}
\caption{The diagram shows the vacuum referred binding energies (VRBE) for electrons of divalent (red) and trivalent (blue) lanthanide ions 
with respect to the vacuum level in BaBrI (a) and BaClI (b). The zigzag red curve with triangles connects the 4f$^{n}$ ground state energies 
of the divalent lanthanide ions. The red curve with circles is attributed to their lowest 4f$^{n-1}$5d$^1$ states. 
Blue zigzag curve with triangles respects to 4f$^n$ ground state energies of trivalent lanthanide ions and blue curve 
with circles are their 4f$^{n-1}$5d$^1$ states. Arrow shows the energy of 4f-5d transition measured in this work. 
The left and right insets provide the calculated density of states (DOS) for BaBrI and BaClI crystals, accordingly.}
\label{dor_diag}
\end{figure*}

Behaviour of spectroscopic properties of the lanthanide ions can be predicted behaviour with respect to their ionic charge and number of electrons into 4f-shell. 
The energy diagrams are plotted for BaBrI and BaClI crystals based on the experimental values of energies of band gap, exciton creation and 4f-5d transition 
and calculated position of ground Eu$^{2+}$ in respect to the valence band obtained in this work.  
The diagram shows the binding energies of an electron in the divalent and trivalent lanthanide ion ground and excited states (Fig.~\ref{dor_diag}). 
By using the Dorenbos chemical shift model \cite{dorenbos2013review},  
these binding energies are related to the binding energy of an electron at rest in vacuum defined as zero of energy 
(Vacuum Referred Binding Energy (VRBE) diagram).

We have not spectrocopic data about trivalent lanthanides in investigated crystals, therefore the Coulomb repulsion energy $U(6,A)$  is used. It defines the binding energy difference between an electron in the ground state of Eu$^{2+}$ with that in the ground state of Eu$^{3+}$ \cite{dorenbos2012modeling, dorenbos2013lanthanide}. Similarly to the other alkali-earth iodide and bromide compounds like LaBr$_3$ and SrI$_2$ it is estimated as 6.4 eV for BaBrI. Chlorides have higher U(6,A) energy, therefore this one is chosen for BaClI equal to 6.9 eV. Such choice immediately defines the VRBE of electrons in the 4f$^n$ states of divalent and trivalent lanthanides. The calculated energies betweeen top of the valence band and Eu$^{2+}$ ground state in BaBrI are about 1.4~eV and 1.5~eV in BaClI crystals (see Table~\ref{all-data}). Energies of valence band top $E_v$ are equal to -5.18 and -5.7~eV for BaBrI and BaClI, respectivelly. Using exciton creation $E_x$ (5.35$\pm$0.15~eV for BaBrI and 6.0$\pm$0.4 eV) and exciton binding $E_b$ (0.23~eV for BaBrI and 0.26~eV for BaClI) energies estimated above the energy of conduction band bottom ($E_c$) is calculated. These obtain about 0.4~eV in BaBrI and 0.53~eV in BaClI. These values are above the vacuum level implying that BaBrI and BaClI should be a negative electron affinity material ($E_C>$0, $\chi=-E_C$). The fact, that $E_C>$0, can be explained by uncertainty in measurements of exciton and exciton binding energies and an error in calculation of distance between Eu$^{2+}$ ground state and top of valence band. It appears that the used in this work \latin{Ab Initio} approach usually underestimates energies of transitions \cite{canning, dudarev}. On the other hand, the chemical shift model gives negative electron affinity $\chi=$-0.6~eV for some materials such as LiCaAlF$_6$, SrAl$_{12}$O$_{19}$ and SrI$_2$ due to the model limitation \cite{dorenbos2013lanthanide,dorenbos2013determining,alekhin2015luminescence}.

Following the diagrams ground states of divalent La, Ce, Gd, and Tb ions are to be 4f5d. 
Stable divalent La, Ce, Gd, and Tb ions and accompanying them photochromic centres as possibly present
as pointed in Refs.~\cite{egranov2016instability, shendrik2016spectroscopy}. 
The divalent state of these ions can be obtained by x-ray irradiation or additive coloration in alkali-earth metal vapours.

It is clear from the diagrams in Fig.~\ref{dor_diag} that all 4f ground levels of trivalent rare earth ions exluded Ce$^{3+}$ lie deeply in valence band. 
Therefore, we expect only 4f-4f luminescence from the trivalent rare earth dopants in these crystals. 
In constrast, the ground 4f and 5d levels of Ce$^{3+}$ ions are in band gap and 5d-4f luminescence in the Ce$^{3+}$ doped crystals can be possible.

The alkali-earth fluorides are not efficient scintillator although the 5d-4f luminescence of Ce$^{3+}$ and Pr$^{3+}$ ions is observed in them. The reason for this is the inefficient energy transfer from hot holes in valence band to rare-earth ions. The energy barrier for hole capturing by the rare earth ion is higher than the energy of hole self-trapping. Therefore, in alkali-earth fluorides energy is transferred through hole traps to rare earth ion. This leads to hyperbolic law of luminescence decay and low light yield \cite{shendrik2012energy}. If the ground state of trivalent rare earth ion is located closer to the top of valence band the probability of hot hole capturing increases. The ground state of Ce$^{3+}$ ion in BaClI and BaBrI crystals as predicted from the chemical shift model is located so close to the top of valence band. Therefore, Ce-doped BaClI and BaBrI crystals would be promising for high light output. 

Notwithstanding, 5d-4f luminescence is possible for trivalent lanthanide ions having ground state lower than top of valence band. Theoretical calculations indicate that valence band of the investigated crystals consists of two subbands formed by p-orbitals of iodine and bromine or chlorine ions (Fig.~\ref{dor_diag},  left and right insets). The iodine valence band lies upper than the bromine and chlorine as well as in BaFBr and BaFCl crystals where valence band is formed by bromine-fluorine and chlorine-fluorine subbands \cite{nicklaus1979optical, ruter1990creation}. Ground state of oxygen-vacancy centres in BaFBr and BaClI crystals is located within valence band in gap between the subbands whereas excited states are located in band gap. Bright luminescence related to the oxygen-vacancy centres was observed \cite{radzhabov1995optical}.  Therefore, if a ground state of lanthanide ion is in gap between the subbands the 5d-4f luminescence is possible. According to the calculations the ground state of Pr$^{3+}$ ion is located into the gap between valence subbands. In this case 5d-4f luminescence is observed in Pr-doped BaBrI and BaClI crystals inasmuch as the lowest 5d state binding energy is less than the one of 4f($^1$S$_0$) level.
The efficient hole capture by the Pr-ions is expected. In Pr-doped alkali-earth fluorides hight temperature stability of light yield was demonstrated in Ref.~\cite{shendrik2010temperature}. We are looking forward to the same effect for Pr-doped BaClI and BaBrI crystals. Thus, based on the predictions of the chemical shift model Ce$^{3+}$ and Pr$^{3+}$ ions seem to be as promising activators for the scintillation BaClI and BaBrI crystals. 

\section{Conclusion}
The scintillation, luminescence and structural properties of BaBrI and BaClI crystals doped with low concentrations of Eu$^{2+}$ ions have been studied. The BaClI and BaBrI single crystal undoped and doped with 0.05-0.1 mol.\% Eu$^{2+}$ ions were grown. The structure of BaClI crystals was determined by single-crystal X-ray diffraction technique. The results obtained by luminescence spectroscopy revealed the presence of divalent europium ions in the BaBrI and BaClI crystals. Although, the impurity ions in the trivalent state in these crystals have not been found. 

The intense bands in absorption and excitation spectra having slightly manifested characteristic "staircase" structure peaked at about 4.25-4.4~eV for BaBrI and 4.25-4.45~eV for BaClI crystals were caused by $\mathrm{4f^{7}(^{8}S_{7/2})} \rightarrow 4f^{6}5d^{1}(t_{2g}) $ transitions. The lowest energies of 4f-5d transition in Eu$^{2+}$ ion obtained from the spectra were 3.29~eV for BaBrI and 3.32~eV for BaClI. The narrow intense peaks at 5.35 and 6~eV in the photoluminescence excitation spectrum of the intrinsic emission of BaBrI and BaClI crystals were due to the creation of excitons. The band gaps of BaBrI and BaClI crystals were estimated about 5.58 and 6.26~eV consequently using hydrogen-like exciton model.

The band gap eas estimated both with the PBE, and $\mathrm{G_{0}W_{0}}$ approximations. The calculated band gap energies agree with the experimental data. The distance between the lowest 5d level of Eu$^{2+}$ ions and top of the valence band has been calculated. The VRBE diagrams of levels of all divalent and trivalent lanthanides ions in BaBrI and BaClI crystals were constructed established on the acquired experimental and theoretical data.

\begin{acknowledgement}

This work was partially supported by RFBR grants 15-02-06514a. The reported study was performed with the equipment set at the Centres for Collective Use ("Isotope-geochemistry investigations" at A.P. Vinogradov Institute of Geochemistry SB RAS and "Analysis of organic state" at A.E. Favorsky Institute of Chemistry SB RAS).
\end{acknowledgement}

\bibliography{photochromic}

\providecommand{\latin}[1]{#1}
\makeatletter
\providecommand{\doi}
  {\begingroup\let\do\@makeother\dospecials
  \catcode`\{=1 \catcode`\}=2\doi@aux}
\providecommand{\doi@aux}[1]{\endgroup\texttt{#1}}
\makeatother
\providecommand*\mcitethebibliography{\thebibliography}
\csname @ifundefined\endcsname{endmcitethebibliography}
  {\let\endmcitethebibliography\endthebibliography}{}
\begin{mcitethebibliography}{60}
\providecommand*\natexlab[1]{#1}
\providecommand*\mciteSetBstSublistMode[1]{}
\providecommand*\mciteSetBstMaxWidthForm[2]{}
\providecommand*\mciteBstWouldAddEndPuncttrue
  {\def\EndOfBibitem{\unskip.}}
\providecommand*\mciteBstWouldAddEndPunctfalse
  {\let\EndOfBibitem\relax}
\providecommand*\mciteSetBstMidEndSepPunct[3]{}
\providecommand*\mciteSetBstSublistLabelBeginEnd[3]{}
\providecommand*\EndOfBibitem{}
\mciteSetBstSublistMode{f}
\mciteSetBstMaxWidthForm{subitem}{(\alph{mcitesubitemcount})}
\mciteSetBstSublistLabelBeginEnd
  {\mcitemaxwidthsubitemform\space}
  {\relax}
  {\relax}

\bibitem[van Loef \latin{et~al.}(2009)van Loef, Wilson, Cherepy, Hull, Payne,
  Choong, Moses, and Shah]{loef2009}
van Loef,~E.~V.; Wilson,~C.~M.; Cherepy,~N.~J.; Hull,~G.; Payne,~S.~A.;
  Choong,~W.-S.; Moses,~W.~W.; Shah,~K.~S. Crystal growth and scintillation
  properties of strontium iodide scintillators. \emph{IEEE Transactions on
  Nuclear Science} \textbf{2009}, \emph{56}, 869\relax
\mciteBstWouldAddEndPuncttrue
\mciteSetBstMidEndSepPunct{\mcitedefaultmidpunct}
{\mcitedefaultendpunct}{\mcitedefaultseppunct}\relax
\EndOfBibitem
\bibitem[Pustovarov \latin{et~al.}(2012)Pustovarov, Ogorodnikov, Goloshumova,
  Isaenko, and Yelisseyev]{pustovarov2012luminescence}
Pustovarov,~V.; Ogorodnikov,~I.; Goloshumova,~A.; Isaenko,~L.; Yelisseyev,~A. A
  luminescence spectroscopy study of scintillation crystals {SrI$_2$} doped
  with {Eu$^{2+}$}. \emph{Optical Materials} \textbf{2012}, \emph{34},
  926--930\relax
\mciteBstWouldAddEndPuncttrue
\mciteSetBstMidEndSepPunct{\mcitedefaultmidpunct}
{\mcitedefaultendpunct}{\mcitedefaultseppunct}\relax
\EndOfBibitem
\bibitem[Pankratov \latin{et~al.}(2013)Pankratov, Popov, Shirmane, Kotlov,
  Bizarri, Burger, Bhattacharya, Tupitsyn, Rowe, and
  Buliga]{pankratov2013luminescence}
Pankratov,~V.; Popov,~A.~I.; Shirmane,~L.; Kotlov,~A.; Bizarri,~G.~A.;
  Burger,~A.; Bhattacharya,~P.; Tupitsyn,~E.; Rowe,~E.; Buliga,~V.~M.
  Luminescence and ultraviolet excitation spectroscopy of {SrI$_2$} and
  {SrI$_{2}$:Eu$^{2+}$}. \emph{Radiation Measurements} \textbf{2013},
  \emph{56}, 13--17\relax
\mciteBstWouldAddEndPuncttrue
\mciteSetBstMidEndSepPunct{\mcitedefaultmidpunct}
{\mcitedefaultendpunct}{\mcitedefaultseppunct}\relax
\EndOfBibitem
\bibitem[Ogorodnikov \latin{et~al.}(2013)Ogorodnikov, Pustovarov, Goloshumova,
  Isaenko, Yelisseyev, and Pashkov]{ogorodnikov2013luminescence}
Ogorodnikov,~I.; Pustovarov,~V.; Goloshumova,~A.; Isaenko,~L.; Yelisseyev,~A.;
  Pashkov,~V. A luminescence spectroscopy study of {SrI$_2$: Nd$^{3+}$} single
  crystals. \emph{Journal of Luminescence} \textbf{2013}, \emph{143},
  101--107\relax
\mciteBstWouldAddEndPuncttrue
\mciteSetBstMidEndSepPunct{\mcitedefaultmidpunct}
{\mcitedefaultendpunct}{\mcitedefaultseppunct}\relax
\EndOfBibitem
\bibitem[Alekhin \latin{et~al.}(2015)Alekhin, Awater, Biner, Kr{\"a}mer,
  de~Haas, and Dorenbos]{alekhin2015luminescence}
Alekhin,~M.~S.; Awater,~R.~H.; Biner,~D.~A.; Kr{\"a}mer,~K.~W.; de~Haas,~J.~T.;
  Dorenbos,~P. Luminescence and spectroscopic properties of {Sm$^{2+}$} and
  {Er$^{3+}$} doped {SrI$_{2}$}. \emph{Journal of luminescence} \textbf{2015},
  \emph{167}, 347--351\relax
\mciteBstWouldAddEndPuncttrue
\mciteSetBstMidEndSepPunct{\mcitedefaultmidpunct}
{\mcitedefaultendpunct}{\mcitedefaultseppunct}\relax
\EndOfBibitem
\bibitem[Bourret-Courchesne \latin{et~al.}(2012)Bourret-Courchesne, Bizarri,
  Borade, Gundiah, Samulon, Yan, and Derenzo]{bourret2012}
Bourret-Courchesne,~E.; Bizarri,~G.; Borade,~R.; Gundiah,~G.; Samulon,~E.;
  Yan,~Z.; Derenzo,~S. Crystal growth and characterization of alkali-earth
  halide scintillators. \emph{Journal of Crystal Growth} \textbf{2012},
  \emph{352}, 78\relax
\mciteBstWouldAddEndPuncttrue
\mciteSetBstMidEndSepPunct{\mcitedefaultmidpunct}
{\mcitedefaultendpunct}{\mcitedefaultseppunct}\relax
\EndOfBibitem
\bibitem[Bourret-Courchesne \latin{et~al.}(2010)Bourret-Courchesne, Bizarri,
  Hanrahan, Gundiah, Yan, and Derenzo]{bourret2010}
Bourret-Courchesne,~E.; Bizarri,~G.; Hanrahan,~S.; Gundiah,~G.; Yan,~Z.;
  Derenzo,~S. $\mathrm{BaBrI:Eu^{2+}}$, a new bright scintillator.
  \emph{Nuclear Instruments and Methods in Physics Research Section A:
  Accelerators, Spectrometers, Detectors and Associated Equipment}
  \textbf{2010}, \emph{613}, 95\relax
\mciteBstWouldAddEndPuncttrue
\mciteSetBstMidEndSepPunct{\mcitedefaultmidpunct}
{\mcitedefaultendpunct}{\mcitedefaultseppunct}\relax
\EndOfBibitem
\bibitem[Gundiah \latin{et~al.}(2010)Gundiah, Bourret-Courchesne, Bizarri,
  Hanrahan, Chaudhry, Canning, Moses, and Derenzo]{gundiah2010}
Gundiah,~G.; Bourret-Courchesne,~E.; Bizarri,~G.; Hanrahan,~S.~M.;
  Chaudhry,~A.; Canning,~A.; Moses,~W.~W.; Derenzo,~S.~E. Scintillation
  Properties of $\mathrm{Eu^{2+}}$ -Activated Barium Fluoroiodide. \emph{IEEE
  Transactions on Nuclear Science} \textbf{2010}, \emph{57}, 1702\relax
\mciteBstWouldAddEndPuncttrue
\mciteSetBstMidEndSepPunct{\mcitedefaultmidpunct}
{\mcitedefaultendpunct}{\mcitedefaultseppunct}\relax
\EndOfBibitem
\bibitem[Bizarri \latin{et~al.}(2011)Bizarri, Bourret-Courchesne, Yan, and
  Derenzo]{bizarri2011scintillation}
Bizarri,~G.; Bourret-Courchesne,~E.~D.; Yan,~Z.; Derenzo,~S.~E. Scintillation
  and Optical Properties of {BaBrI:Eu$^{2+}$} and {CsBa$_2$I:Eu$^{2+}$}.
  \emph{IEEE Transactions on Nuclear Science} \textbf{2011}, \emph{58},
  3403--3410\relax
\mciteBstWouldAddEndPuncttrue
\mciteSetBstMidEndSepPunct{\mcitedefaultmidpunct}
{\mcitedefaultendpunct}{\mcitedefaultseppunct}\relax
\EndOfBibitem
\bibitem[Gundiah \latin{et~al.}(2011)Gundiah, Bizarri, Hanrahan, Weber,
  Bourret-Courchesne, and Derenzo]{gundiah2011structure}
Gundiah,~G.; Bizarri,~G.; Hanrahan,~S.~M.; Weber,~M.~J.;
  Bourret-Courchesne,~E.~D.; Derenzo,~S.~E. Structure and scintillation of
  {Eu$^{2+}$}-activated solid solutions in the {BaBr$_2$}--{BaI$_2$} system.
  \emph{Nuclear Instruments and Methods in Physics Research Section A:
  Accelerators, Spectrometers, Detectors and Associated Equipment}
  \textbf{2011}, \emph{652}, 234--237\relax
\mciteBstWouldAddEndPuncttrue
\mciteSetBstMidEndSepPunct{\mcitedefaultmidpunct}
{\mcitedefaultendpunct}{\mcitedefaultseppunct}\relax
\EndOfBibitem
\bibitem[Yan \latin{et~al.}(2016)Yan, Shalapska, and
  Bourret]{yan2016czochralski}
Yan,~Z.; Shalapska,~T.; Bourret,~E. Czochralski growth of the mixed halides
  {BaBrCl} and {BaBrCl:Eu}. \emph{Journal of Crystal Growth} \textbf{2016},
  \emph{435}, 42--45\relax
\mciteBstWouldAddEndPuncttrue
\mciteSetBstMidEndSepPunct{\mcitedefaultmidpunct}
{\mcitedefaultendpunct}{\mcitedefaultseppunct}\relax
\EndOfBibitem
\bibitem[Bruker(2003)]{bruker_2003}
Bruker, {APEX2}. \emph{{Bruker AXS Inc.}, {Madison}, {Wisconsin}, {USA}}
  \textbf{2003}, \relax
\mciteBstWouldAddEndPunctfalse
\mciteSetBstMidEndSepPunct{\mcitedefaultmidpunct}
{}{\mcitedefaultseppunct}\relax
\EndOfBibitem
\bibitem[Bruker(2007)]{bruker_2007}
Bruker, {SAINT}. \emph{{Bruker AXS Inc.}, {Madison}, {Wisconsin}, {USA}}
  \textbf{2007}, \relax
\mciteBstWouldAddEndPunctfalse
\mciteSetBstMidEndSepPunct{\mcitedefaultmidpunct}
{}{\mcitedefaultseppunct}\relax
\EndOfBibitem
\bibitem[Sheldrick(2003)]{sheldrik}
Sheldrick,~G. SADABS, Program for Empirical Absorption Correction of Area
  Detector Data;. \emph{University of G{\"o}ttingen: G{\"o}ttingen, Germany}
  \textbf{2003}, \relax
\mciteBstWouldAddEndPunctfalse
\mciteSetBstMidEndSepPunct{\mcitedefaultmidpunct}
{}{\mcitedefaultseppunct}\relax
\EndOfBibitem
\bibitem[Betteridge \latin{et~al.}(2003)Betteridge, Carruthers, Cooper, Prout,
  and Watkin]{betteridge}
Betteridge,~P.~W.; Carruthers,~J.~R.; Cooper,~R.~I.; Prout,~K.; Watkin,~D.~J.
  {CRYSTALS} version 12: software for guided crystal structure analysis.
  \emph{Journal of Applied Crystallography} \textbf{2003}, \emph{36},
  1487--1487\relax
\mciteBstWouldAddEndPuncttrue
\mciteSetBstMidEndSepPunct{\mcitedefaultmidpunct}
{\mcitedefaultendpunct}{\mcitedefaultseppunct}\relax
\EndOfBibitem
\bibitem[Palatinus and Chapuis(2007)Palatinus, and Chapuis]{palatinus}
Palatinus,~L.; Chapuis,~G. Superflip--a computer program for the solution of
  crystal structures by charge flipping in arbitrary dimensions. \emph{Journal
  of Applied Crystallography} \textbf{2007}, \emph{40}, 786--790\relax
\mciteBstWouldAddEndPuncttrue
\mciteSetBstMidEndSepPunct{\mcitedefaultmidpunct}
{\mcitedefaultendpunct}{\mcitedefaultseppunct}\relax
\EndOfBibitem
\bibitem[Lenus \latin{et~al.}(2002)Lenus, Sornadurai, Rajan, and
  Purniah]{lenus2002luminescence}
Lenus,~A.~J.; Sornadurai,~D.; Rajan,~K.~G.; Purniah,~B. Luminescence behaviour
  of {Eu$^{2+}$}-doped BaClI and BaBrI. \emph{Materials Letters} \textbf{2002},
  \emph{57}, 635--638\relax
\mciteBstWouldAddEndPuncttrue
\mciteSetBstMidEndSepPunct{\mcitedefaultmidpunct}
{\mcitedefaultendpunct}{\mcitedefaultseppunct}\relax
\EndOfBibitem
\bibitem[Kresse and Hafner(1993)Kresse, and Hafner]{vasp}
Kresse,~G.; Hafner,~J. Ab initio molecular dynamics for liquid metals.
  \emph{Physical Review B} \textbf{1993}, \emph{47}, 558\relax
\mciteBstWouldAddEndPuncttrue
\mciteSetBstMidEndSepPunct{\mcitedefaultmidpunct}
{\mcitedefaultendpunct}{\mcitedefaultseppunct}\relax
\EndOfBibitem
\bibitem[mat()]{matrosov}
\emph{Irkutsk Supercomputer Centre of SB RAS http://hpc.icc.ru} \relax
\mciteBstWouldAddEndPunctfalse
\mciteSetBstMidEndSepPunct{\mcitedefaultmidpunct}
{}{\mcitedefaultseppunct}\relax
\EndOfBibitem
\bibitem[foc()]{fock}
\emph{{A.M. Fok} computational cluster of Irkutsk National Researcher Technical
  University http://fock.istu.edu} \relax
\mciteBstWouldAddEndPunctfalse
\mciteSetBstMidEndSepPunct{\mcitedefaultmidpunct}
{}{\mcitedefaultseppunct}\relax
\EndOfBibitem
\bibitem[Perdew \latin{et~al.}(1996)Perdew, Burke, and Ernzerhof]{pbe}
Perdew,~J.~P.; Burke,~K.; Ernzerhof,~M. Generalized gradient approximation made
  simple. \emph{Physical review letters} \textbf{1996}, \emph{77}, 3865\relax
\mciteBstWouldAddEndPuncttrue
\mciteSetBstMidEndSepPunct{\mcitedefaultmidpunct}
{\mcitedefaultendpunct}{\mcitedefaultseppunct}\relax
\EndOfBibitem
\bibitem[Dorenbos(2003)]{dorenbos2003energy}
Dorenbos,~P. Energy of the first {4f$^7$$\rightarrow$4f$^6$5d} transition of
  {$Eu^{2+}$} in inorganic compounds. \emph{Journal of luminescence}
  \textbf{2003}, \emph{104}, 239--260\relax
\mciteBstWouldAddEndPuncttrue
\mciteSetBstMidEndSepPunct{\mcitedefaultmidpunct}
{\mcitedefaultendpunct}{\mcitedefaultseppunct}\relax
\EndOfBibitem
\bibitem[Kobayasi \latin{et~al.}(1980)Kobayasi, Mroczkowski, Owen, and
  Brixner]{kobayasi1980fluorescence}
Kobayasi,~T.; Mroczkowski,~S.; Owen,~J.~F.; Brixner,~L.~H. Fluorescence
  lifetime and quantum efficiency for {5d$\rightarrow$4f} transitions in
  {Eu$^{2+}$} doped chloride and fluoride crystals. \emph{Journal of
  Luminescence} \textbf{1980}, \emph{21}, 247--257\relax
\mciteBstWouldAddEndPuncttrue
\mciteSetBstMidEndSepPunct{\mcitedefaultmidpunct}
{\mcitedefaultendpunct}{\mcitedefaultseppunct}\relax
\EndOfBibitem
\bibitem[Chaudhry \latin{et~al.}(2014)Chaudhry, Boutchko, Chourou, Zhang,
  Gr{\o}nbech-Jensen, and Canning]{chaudhly}
Chaudhry,~A.; Boutchko,~R.; Chourou,~S.; Zhang,~G.; Gr{\o}nbech-Jensen,~N.;
  Canning,~A. First-principles study of luminescence in {Eu$^{2+}$}-doped
  inorganic scintillators. \emph{Physical Review B} \textbf{2014}, \emph{89},
  155105\relax
\mciteBstWouldAddEndPuncttrue
\mciteSetBstMidEndSepPunct{\mcitedefaultmidpunct}
{\mcitedefaultendpunct}{\mcitedefaultseppunct}\relax
\EndOfBibitem
\bibitem[Shendrik \latin{et~al.}(2013)Shendrik, Radzhabov, and
  Nepomnyashchikh]{shendrik2013scintillation}
Shendrik,~R.; Radzhabov,~E.; Nepomnyashchikh,~A. Scintillation properties of
  pure and {Ce$^{3+}$}-doped {SrF$_2$} crystals. \emph{Radiation Measurements}
  \textbf{2013}, \emph{56}, 58--61\relax
\mciteBstWouldAddEndPuncttrue
\mciteSetBstMidEndSepPunct{\mcitedefaultmidpunct}
{\mcitedefaultendpunct}{\mcitedefaultseppunct}\relax
\EndOfBibitem
\bibitem[Shendrik and Radzhabov(2014)Shendrik, and
  Radzhabov]{shendrik2014absolute}
Shendrik,~R.; Radzhabov,~E. Absolute Light Yield Measurements on {SrF$_2$} and
  {BaF$_2$} Doped With Rare Earth Ions. \emph{IEEE Transactions on Nuclear
  Science} \textbf{2014}, \emph{61}, 406--410\relax
\mciteBstWouldAddEndPuncttrue
\mciteSetBstMidEndSepPunct{\mcitedefaultmidpunct}
{\mcitedefaultendpunct}{\mcitedefaultseppunct}\relax
\EndOfBibitem
\bibitem[Bourret-Courchesne \latin{et~al.}(2012)Bourret-Courchesne, Bizarri,
  Borade, Gundiah, Samulon, Yan, and Derenzo]{BourretCourchesne}
Bourret-Courchesne,~E.; Bizarri,~G.; Borade,~R.; Gundiah,~G.; Samulon,~E.;
  Yan,~Z.; Derenzo,~S. Crystal growth and characterization of alkali-earth
  halide scintillators. \emph{Journal of Crystal Growth} \textbf{2012},
  \emph{352}, 78 -- 83\relax
\mciteBstWouldAddEndPuncttrue
\mciteSetBstMidEndSepPunct{\mcitedefaultmidpunct}
{\mcitedefaultendpunct}{\mcitedefaultseppunct}\relax
\EndOfBibitem
\bibitem[Shendrik and Radzhabov(2012)Shendrik, and
  Radzhabov]{shendrik2012energy}
Shendrik,~R.; Radzhabov,~E. Energy Transfer Mechanism in {Pr}-Doped {SrF$_2$}
  Crystals. \emph{IEEE transactions on nuclear science} \textbf{2012},
  \emph{59}, 2089--2094\relax
\mciteBstWouldAddEndPuncttrue
\mciteSetBstMidEndSepPunct{\mcitedefaultmidpunct}
{\mcitedefaultendpunct}{\mcitedefaultseppunct}\relax
\EndOfBibitem
\bibitem[Song and Williams(2013)Song, and Williams]{song2013self}
Song,~K.; Williams,~R.~T. \emph{Self-trapped excitons}; Springer Science \&
  Business Media, 2013; Vol. 105\relax
\mciteBstWouldAddEndPuncttrue
\mciteSetBstMidEndSepPunct{\mcitedefaultmidpunct}
{\mcitedefaultendpunct}{\mcitedefaultseppunct}\relax
\EndOfBibitem
\bibitem[Beaumont \latin{et~al.}(1970)Beaumont, Hayes, Kirk, and
  Summers]{beaumont1970investigation}
Beaumont,~J.; Hayes,~W.; Kirk,~D.; Summers,~G. An investigation of trapped
  holes and trapped excitons in alkaline earth fluorides. Proceedings of the
  Royal Society of London A: Mathematical, Physical and Engineering Sciences.
  1970; pp 69--97\relax
\mciteBstWouldAddEndPuncttrue
\mciteSetBstMidEndSepPunct{\mcitedefaultmidpunct}
{\mcitedefaultendpunct}{\mcitedefaultseppunct}\relax
\EndOfBibitem
\bibitem[Radzhabov and Egranov(1994)Radzhabov, and
  Egranov]{radzhabov1994exciton}
Radzhabov,~E.; Egranov,~A. Exciton emission in {BaFBr} and {BaFCl} crystals.
  \emph{Journal of Physics: Condensed Matter} \textbf{1994}, \emph{6},
  5639\relax
\mciteBstWouldAddEndPuncttrue
\mciteSetBstMidEndSepPunct{\mcitedefaultmidpunct}
{\mcitedefaultendpunct}{\mcitedefaultseppunct}\relax
\EndOfBibitem
\bibitem[Radzhabov and Otroshok(1995)Radzhabov, and
  Otroshok]{radzhabov1995optical}
Radzhabov,~E.; Otroshok,~V. Optical spectra of oxygen defects in {BaFCl} and
  {BaFBr} crystals. \emph{Journal of Physics and Chemistry of Solids}
  \textbf{1995}, \emph{56}, 1--7\relax
\mciteBstWouldAddEndPuncttrue
\mciteSetBstMidEndSepPunct{\mcitedefaultmidpunct}
{\mcitedefaultendpunct}{\mcitedefaultseppunct}\relax
\EndOfBibitem
\bibitem[Radzhabov(1994)]{radzhabov1994time}
Radzhabov,~E. Time-resolved luminescence of oxygen-vacancy centres in
  alkaline-earth fluoride and barium fluorohalide crystals. \emph{Journal of
  Physics: Condensed Matter} \textbf{1994}, \emph{6}, 9807\relax
\mciteBstWouldAddEndPuncttrue
\mciteSetBstMidEndSepPunct{\mcitedefaultmidpunct}
{\mcitedefaultendpunct}{\mcitedefaultseppunct}\relax
\EndOfBibitem
\bibitem[Dvorak \latin{et~al.}(2013)Dvorak, Wei, and Wu]{dvorak2013origin}
Dvorak,~M.; Wei,~S.-H.; Wu,~Z. Origin of the variation of exciton binding
  energy in semiconductors. \emph{Physical review letters} \textbf{2013},
  \emph{110}, 016402\relax
\mciteBstWouldAddEndPuncttrue
\mciteSetBstMidEndSepPunct{\mcitedefaultmidpunct}
{\mcitedefaultendpunct}{\mcitedefaultseppunct}\relax
\EndOfBibitem
\bibitem[Ramamurti and Teegarden(1966)Ramamurti, and
  Teegarden]{ramamurti1966intrinsic}
Ramamurti,~J.; Teegarden,~K. Intrinsic Luminescence of {RbI} and {KI} at
  {10$^{\circ}$ K}. \emph{Physical Review} \textbf{1966}, \emph{145}, 698\relax
\mciteBstWouldAddEndPuncttrue
\mciteSetBstMidEndSepPunct{\mcitedefaultmidpunct}
{\mcitedefaultendpunct}{\mcitedefaultseppunct}\relax
\EndOfBibitem
\bibitem[Williams and Kabler(1974)Williams, and Kabler]{williams1974excited}
Williams,~R.; Kabler,~M. Excited-state absorption spectroscopy of self-trapped
  excitons in alkali halides. \emph{Physical Review B} \textbf{1974}, \emph{9},
  1897\relax
\mciteBstWouldAddEndPuncttrue
\mciteSetBstMidEndSepPunct{\mcitedefaultmidpunct}
{\mcitedefaultendpunct}{\mcitedefaultseppunct}\relax
\EndOfBibitem
\bibitem[Onodera and Toyozawa(1967)Onodera, and Toyozawa]{onodera1967excitons}
Onodera,~Y.; Toyozawa,~Y. Excitons in alkali halides. \emph{Journal of the
  Physical Society of Japan} \textbf{1967}, \emph{22}, 833--844\relax
\mciteBstWouldAddEndPuncttrue
\mciteSetBstMidEndSepPunct{\mcitedefaultmidpunct}
{\mcitedefaultendpunct}{\mcitedefaultseppunct}\relax
\EndOfBibitem
\bibitem[Tomiki and Miyata(1969)Tomiki, and Miyata]{tomiki1969optical}
Tomiki,~T.; Miyata,~T. Optical studies of alkali fluorides and alkaline earth
  fluorides in {VUV} region. \emph{Journal of the Physical Society of Japan}
  \textbf{1969}, \emph{27}, 658--678\relax
\mciteBstWouldAddEndPuncttrue
\mciteSetBstMidEndSepPunct{\mcitedefaultmidpunct}
{\mcitedefaultendpunct}{\mcitedefaultseppunct}\relax
\EndOfBibitem
\bibitem[Nicklaus(1979)]{nicklaus1979optical}
Nicklaus,~E. Optical properties of some alkaline earth halides. \emph{physica
  status solidi (a)} \textbf{1979}, \emph{53}, 217--224\relax
\mciteBstWouldAddEndPuncttrue
\mciteSetBstMidEndSepPunct{\mcitedefaultmidpunct}
{\mcitedefaultendpunct}{\mcitedefaultseppunct}\relax
\EndOfBibitem
\bibitem[Ayachour \latin{et~al.}(1991)Ayachour, Sieskind, and
  Geist]{ayachour1991electrical}
Ayachour,~D.; Sieskind,~M.; Geist,~P. Electrical properties of alkaline earth
  fluorohalide crystals. \emph{physica status solidi (b)} \textbf{1991},
  \emph{166}, 43--52\relax
\mciteBstWouldAddEndPuncttrue
\mciteSetBstMidEndSepPunct{\mcitedefaultmidpunct}
{\mcitedefaultendpunct}{\mcitedefaultseppunct}\relax
\EndOfBibitem
\bibitem[Tsujibayashi \latin{et~al.}(1999)Tsujibayashi, Watanabe, Arimoto,
  Itoh, Nakanishi, Itoh, Asaka, and Kamada]{tsujibayashi1999resonant}
Tsujibayashi,~T.; Watanabe,~M.; Arimoto,~O.; Itoh,~M.; Nakanishi,~S.; Itoh,~H.;
  Asaka,~S.; Kamada,~M. Resonant enhancement effect on two-photon absorption
  due to excitons in alkaline-earth fluorides excited with synchrotron
  radiation and laser light. \emph{Physical Review B} \textbf{1999}, \emph{60},
  R8442\relax
\mciteBstWouldAddEndPuncttrue
\mciteSetBstMidEndSepPunct{\mcitedefaultmidpunct}
{\mcitedefaultendpunct}{\mcitedefaultseppunct}\relax
\EndOfBibitem
\bibitem[Belsky and Krupa(1999)Belsky, and Krupa]{belsky1999luminescence}
Belsky,~A.; Krupa,~J. Luminescence excitation mechanisms of rare earth doped
  phosphors in the {VUV} range. \emph{Displays} \textbf{1999}, \emph{19},
  185--196\relax
\mciteBstWouldAddEndPuncttrue
\mciteSetBstMidEndSepPunct{\mcitedefaultmidpunct}
{\mcitedefaultendpunct}{\mcitedefaultseppunct}\relax
\EndOfBibitem
\bibitem[Dorenbos(2005)]{dorenbos2005eu3+}
Dorenbos,~P. The {Eu$^{3+}$} charge transfer energy and the relation with the
  band gap of compounds. \emph{Journal of luminescence} \textbf{2005},
  \emph{111}, 89--104\relax
\mciteBstWouldAddEndPuncttrue
\mciteSetBstMidEndSepPunct{\mcitedefaultmidpunct}
{\mcitedefaultendpunct}{\mcitedefaultseppunct}\relax
\EndOfBibitem
\bibitem[Joos \latin{et~al.}(2015)Joos, Poelman, and Smet]{joos2015energy}
Joos,~J.~J.; Poelman,~D.; Smet,~P.~F. Energy level modeling of lanthanide
  materials: review and uncertainty analysis. \emph{Physical Chemistry Chemical
  Physics} \textbf{2015}, \emph{17}, 19058--19078\relax
\mciteBstWouldAddEndPuncttrue
\mciteSetBstMidEndSepPunct{\mcitedefaultmidpunct}
{\mcitedefaultendpunct}{\mcitedefaultseppunct}\relax
\EndOfBibitem
\bibitem[Canning \latin{et~al.}(2011)Canning, Chaudhry, Boutchko, and
  Gr{\o}nbech-Jensen]{canning}
Canning,~A.; Chaudhry,~A.; Boutchko,~R.; Gr{\o}nbech-Jensen,~N.
  First-principles study of luminescence in {Ce}-doped inorganic scintillators.
  \emph{Physical Review B} \textbf{2011}, \emph{83}, 125115\relax
\mciteBstWouldAddEndPuncttrue
\mciteSetBstMidEndSepPunct{\mcitedefaultmidpunct}
{\mcitedefaultendpunct}{\mcitedefaultseppunct}\relax
\EndOfBibitem
\bibitem[Dudarev \latin{et~al.}(1998)Dudarev, Botton, Savrasov, Humphreys, and
  Sutton]{dudarev}
Dudarev,~S.; Botton,~G.; Savrasov,~S.; Humphreys,~C.; Sutton,~A.
  Electron-energy-loss spectra and the structural stability of nickel oxide: An
  {LSDA+U} study. \emph{Physical Review B} \textbf{1998}, \emph{57}, 1505\relax
\mciteBstWouldAddEndPuncttrue
\mciteSetBstMidEndSepPunct{\mcitedefaultmidpunct}
{\mcitedefaultendpunct}{\mcitedefaultseppunct}\relax
\EndOfBibitem
\bibitem[H{\"o}ls{\"a} \latin{et~al.}(2009)H{\"o}ls{\"a}, Kirm, Laamanen,
  Lastusaari, Niittykoski, and Nov{\'a}k]{holsa}
H{\"o}ls{\"a},~J.; Kirm,~M.; Laamanen,~T.; Lastusaari,~M.; Niittykoski,~J.;
  Nov{\'a}k,~P. Electronic structure of the {Sr$_2$MgSi$_2$O$_7$: Eu$^{2+}$}
  persistent luminescence material. \emph{Journal of Luminescence}
  \textbf{2009}, \emph{129}, 1560--1563\relax
\mciteBstWouldAddEndPuncttrue
\mciteSetBstMidEndSepPunct{\mcitedefaultmidpunct}
{\mcitedefaultendpunct}{\mcitedefaultseppunct}\relax
\EndOfBibitem
\bibitem[Shi \latin{et~al.}(2008)Shi, Ouyang, Fang, Shen, Tang, and Li]{shi}
Shi,~S.; Ouyang,~C.; Fang,~Q.; Shen,~J.; Tang,~W.; Li,~C. Electronic structure
  and magnetism of EuX (X= O, S, Se and Te): A first-principles investigation.
  \emph{EPL (Europhysics Letters)} \textbf{2008}, \emph{83}, 69001\relax
\mciteBstWouldAddEndPuncttrue
\mciteSetBstMidEndSepPunct{\mcitedefaultmidpunct}
{\mcitedefaultendpunct}{\mcitedefaultseppunct}\relax
\EndOfBibitem
\bibitem[Shishkin and Kresse(2007)Shishkin, and Kresse]{shishkin}
Shishkin,~M.; Kresse,~G. Self-consistent G W calculations for semiconductors
  and insulators. \emph{Physical Review B} \textbf{2007}, \emph{75},
  235102\relax
\mciteBstWouldAddEndPuncttrue
\mciteSetBstMidEndSepPunct{\mcitedefaultmidpunct}
{\mcitedefaultendpunct}{\mcitedefaultseppunct}\relax
\EndOfBibitem
\bibitem[Fuchs \latin{et~al.}(2007)Fuchs, Furthm{\"u}ller, Bechstedt, Shishkin,
  and Kresse]{fuchs}
Fuchs,~F.; Furthm{\"u}ller,~J.; Bechstedt,~F.; Shishkin,~M.; Kresse,~G.
  Quasiparticle band structure based on a generalized Kohn-Sham scheme.
  \emph{Physical Review B} \textbf{2007}, \emph{76}, 115109\relax
\mciteBstWouldAddEndPuncttrue
\mciteSetBstMidEndSepPunct{\mcitedefaultmidpunct}
{\mcitedefaultendpunct}{\mcitedefaultseppunct}\relax
\EndOfBibitem
\bibitem[Lee \latin{et~al.}(2016)Lee, Seko, Shitara, Nakayama, and Tanaka]{lee}
Lee,~J.; Seko,~A.; Shitara,~K.; Nakayama,~K.; Tanaka,~I. Prediction model of
  band gap for inorganic compounds by combination of density functional theory
  calculations and machine learning techniques. \emph{Physical Review B}
  \textbf{2016}, \emph{93}, 115104\relax
\mciteBstWouldAddEndPuncttrue
\mciteSetBstMidEndSepPunct{\mcitedefaultmidpunct}
{\mcitedefaultendpunct}{\mcitedefaultseppunct}\relax
\EndOfBibitem
\bibitem[Dorenbos(2013)]{dorenbos2013review}
Dorenbos,~P. A review on how lanthanide impurity levels change with chemistry
  and structure of inorganic compounds. \emph{ECS Journal of Solid State
  Science and Technology} \textbf{2013}, \emph{2}, R3001--R3011\relax
\mciteBstWouldAddEndPuncttrue
\mciteSetBstMidEndSepPunct{\mcitedefaultmidpunct}
{\mcitedefaultendpunct}{\mcitedefaultseppunct}\relax
\EndOfBibitem
\bibitem[Dorenbos(2012)]{dorenbos2012modeling}
Dorenbos,~P. Modeling the chemical shift of lanthanide 4f electron binding
  energies. \emph{Physical Review B} \textbf{2012}, \emph{85}, 165107\relax
\mciteBstWouldAddEndPuncttrue
\mciteSetBstMidEndSepPunct{\mcitedefaultmidpunct}
{\mcitedefaultendpunct}{\mcitedefaultseppunct}\relax
\EndOfBibitem
\bibitem[Dorenbos(2013)]{dorenbos2013lanthanide}
Dorenbos,~P. Lanthanide 4f-electron binding energies and the nephelauxetic
  effect in wide band gap compounds. \emph{Journal of Luminescence}
  \textbf{2013}, \emph{136}, 122--129\relax
\mciteBstWouldAddEndPuncttrue
\mciteSetBstMidEndSepPunct{\mcitedefaultmidpunct}
{\mcitedefaultendpunct}{\mcitedefaultseppunct}\relax
\EndOfBibitem
\bibitem[Dorenbos(2013)]{dorenbos2013determining}
Dorenbos,~P. Determining binding energies of valence-band electrons in
  insulators and semiconductors via lanthanide spectroscopy. \emph{Physical
  Review B} \textbf{2013}, \emph{87}, 035118\relax
\mciteBstWouldAddEndPuncttrue
\mciteSetBstMidEndSepPunct{\mcitedefaultmidpunct}
{\mcitedefaultendpunct}{\mcitedefaultseppunct}\relax
\EndOfBibitem
\bibitem[Egranov \latin{et~al.}(2016)Egranov, Sizova, Shendrik, and
  Smirnova]{egranov2016instability}
Egranov,~A.; Sizova,~T.~Y.; Shendrik,~R.~Y.; Smirnova,~N. Instability of some
  divalent rare earth ions and photochromic effect. \emph{Journal of Physics
  and Chemistry of Solids} \textbf{2016}, \emph{90}, 7--15\relax
\mciteBstWouldAddEndPuncttrue
\mciteSetBstMidEndSepPunct{\mcitedefaultmidpunct}
{\mcitedefaultendpunct}{\mcitedefaultseppunct}\relax
\EndOfBibitem
\bibitem[Shendrik \latin{et~al.}(2016)Shendrik, Myasnikova, Radzhabov, and
  Nepomnyashchikh]{shendrik2016spectroscopy}
Shendrik,~R.; Myasnikova,~A.; Radzhabov,~E.; Nepomnyashchikh,~A. Spectroscopy
  of divalent rare earth ions in fluoride crystals. \emph{Journal of
  Luminescence} \textbf{2016}, \emph{169}, 635--640\relax
\mciteBstWouldAddEndPuncttrue
\mciteSetBstMidEndSepPunct{\mcitedefaultmidpunct}
{\mcitedefaultendpunct}{\mcitedefaultseppunct}\relax
\EndOfBibitem
\bibitem[R{\"u}ter \latin{et~al.}(1990)R{\"u}ter, Seggern, Reininger, and
  Saile]{ruter1990creation}
R{\"u}ter,~H.; Seggern,~H.~v.; Reininger,~R.; Saile,~V. Creation of
  photostimulable centers in BaFBr:{Eu$^{2+}$} single crystals by vacuum
  ultraviolet radiation. \emph{Physical review letters} \textbf{1990},
  \emph{65}, 2438\relax
\mciteBstWouldAddEndPuncttrue
\mciteSetBstMidEndSepPunct{\mcitedefaultmidpunct}
{\mcitedefaultendpunct}{\mcitedefaultseppunct}\relax
\EndOfBibitem
\bibitem[Shendrik and Radzhabov(2010)Shendrik, and
  Radzhabov]{shendrik2010temperature}
Shendrik,~R.; Radzhabov,~E. Temperature Dependence of {Ce$^{3+}$} and
  {Pr$^{3+}$} Emission in {CaF$_2$}, {SrF$_2$}, and {BaF$_2$}. \emph{IEEE
  Transactions on Nuclear science} \textbf{2010}, \emph{57}, 1295--1299\relax
\mciteBstWouldAddEndPuncttrue
\mciteSetBstMidEndSepPunct{\mcitedefaultmidpunct}
{\mcitedefaultendpunct}{\mcitedefaultseppunct}\relax
\EndOfBibitem
\end{mcitethebibliography}

\end{document}